\documentclass{article}

\usepackage{microtype}
\usepackage{graphicx}
\usepackage{subfigure}
\usepackage{booktabs} 
\usepackage{caption}

\newcommand{\ours}{MolPilot}

\newcounter{mycounter}

\newcommand{\ul}[1]{{\underline{#1}}}

\usepackage{microtype}
\usepackage{graphicx}
\usepackage{subfigure}
\usepackage{booktabs} 
\usepackage{paralist}
\usepackage{tablefootnote}
\usepackage{multirow, multicol}

\usepackage{amsmath,amsfonts,bm}

\def\eqref#1{equation~\ref{#1}}

\def\1{\bm{1}}

\def\eps{{\epsilon}}

\def\rve{{\mathbf{e}}}

\def\rvx{{\mathbf{x}}}
\def\rvy{{\mathbf{y}}}

\def\rmF{{\mathbf{F}}}

\def\vtheta{{\bm{\theta}}}

\def\vh{{\bm{h}}}

\def\vr{{\bm{r}}}

\def\vx{{\bm{x}}}

\def\mA{{\bm{A}}}

\def\mC{{\bm{C}}}

\def\mI{{\bm{I}}}

\def\mW{{\bm{W}}}
\def\mX{{\bm{X}}}

\DeclareMathAlphabet{\mathsfit}{\encodingdefault}{\sfdefault}{m}{sl}
\SetMathAlphabet{\mathsfit}{bold}{\encodingdefault}{\sfdefault}{bx}{n}

\def\gL{{\mathcal{L}}}

\def\gN{{\mathcal{N}}}

\def\gX{{\mathcal{X}}}

\def\gZ{{\mathcal{Z}}}

\newcommand{\E}{\mathbb{E}}

\newcommand{\R}{\mathbb{R}}

\newcommand{\KL}{D_{\mathrm{KL}}}

\DeclareMathOperator*{\argmin}{arg\,min}

\usepackage[ruled, vlined, boxed]{algorithm2e}
\usepackage[table]{xcolor}
\usepackage{hyperref}

\usepackage[accepted]{icml2025}

\usepackage{amsmath}
\usepackage{amssymb}
\usepackage{mathtools}
\usepackage{amsthm}

\usepackage[capitalize,noabbrev]{cleveref}

\theoremstyle{plain}
\newtheorem{theorem}{Theorem}[section]
\newtheorem{proposition}[theorem]{Proposition}

\theoremstyle{definition}
\newtheorem{definition}[theorem]{Definition}

\theoremstyle{remark}
\newtheorem{remark}[theorem]{Remark}

\usepackage[textsize=tiny]{todonotes}

\icmltitlerunning{Piloting Structure-Based Drug Design via Modality-Specific Optimal Schedule}

\begin{document}
\definecolor{mygreen}{HTML}{00bc12}

\twocolumn[
\icmltitle{Piloting Structure-Based Drug Design via Modality-Specific Optimal Schedule}



\icmlsetsymbol{equal}{*}

\begin{icmlauthorlist}
\icmlauthor{Keyue Qiu}{equal,air,thu}
\icmlauthor{Yuxuan Song}{equal,air,thu}
\icmlauthor{Zhehuan Fan}{shanghai}
\icmlauthor{Peidong Liu}{sichuan} \\
\icmlauthor{Zhe Zhang}{thu}
\icmlauthor{Mingyue Zheng}{shanghai}
\icmlauthor{Hao Zhou}{air}
\icmlauthor{Wei-Ying Ma}{air}
\end{icmlauthorlist}

\icmlaffiliation{thu}{Department of Computer Science and Technology, Tsinghua University}
\icmlaffiliation{air}{Institute for AI Industry Research (AIR), Tsinghua University}
\icmlaffiliation{shanghai}{Shanghai Institute of Materia Medica, Chinese Academy of Sciences}
\icmlaffiliation{sichuan}{Sichuan University}

\icmlcorrespondingauthor{Hao Zhou}{zhouhao@air.tsinghua.edu.cn}

\icmlkeywords{Optimal Scheduling, Bayesian Flow Network, Structure-based Drug Design}

\vskip 0.3in
]



\printAffiliationsAndNotice{\icmlEqualContribution} 

\begin{abstract}
Structure-Based Drug Design (SBDD) is crucial for identifying bioactive molecules. Recent deep generative models are faced with challenges in geometric structure modeling. A major bottleneck lies in the twisted probability path of multi-modalities—continuous 3D positions and discrete 2D topologies—which jointly determine molecular geometries. By establishing the fact that noise schedules decide the Variational Lower Bound (VLB) for the twisted probability path, we propose VLB-Optimal Scheduling (VOS) strategy in this under-explored area, which optimizes VLB as a path integral for SBDD. Our model effectively enhances molecular geometries and interaction modeling, achieving a state-of-the-art PoseBusters passing rate of 95.9\% on CrossDock, more than 10\% improvement upon strong baselines, while maintaining high affinities and robust intramolecular validity evaluated on a held-out test set.
Code is available at \url{https://github.com/AlgoMole/MolCRAFT}.
\end{abstract}

\section{Introduction}

Structure-Based Drug Design (SBDD) plays a pivotal role in the discovery of bioactive molecules, leveraging the knowledge of protein-ligand interactions to identify potential therapeutic compounds. At the core of SBDD is the accurate modeling of 3D protein-ligand geometries, as only when bioactive compounds can bind effectively to their target receptors can they elicit their therapeutic effects \citep{isert_structure-based_2023}. Despite its importance, achieving high-fidelity interaction modeling remains a significant challenge, primarily due to the complexity of the underlying binding dynamics.

\begin{figure}[t]
    \centering
    \includegraphics[width=\linewidth]{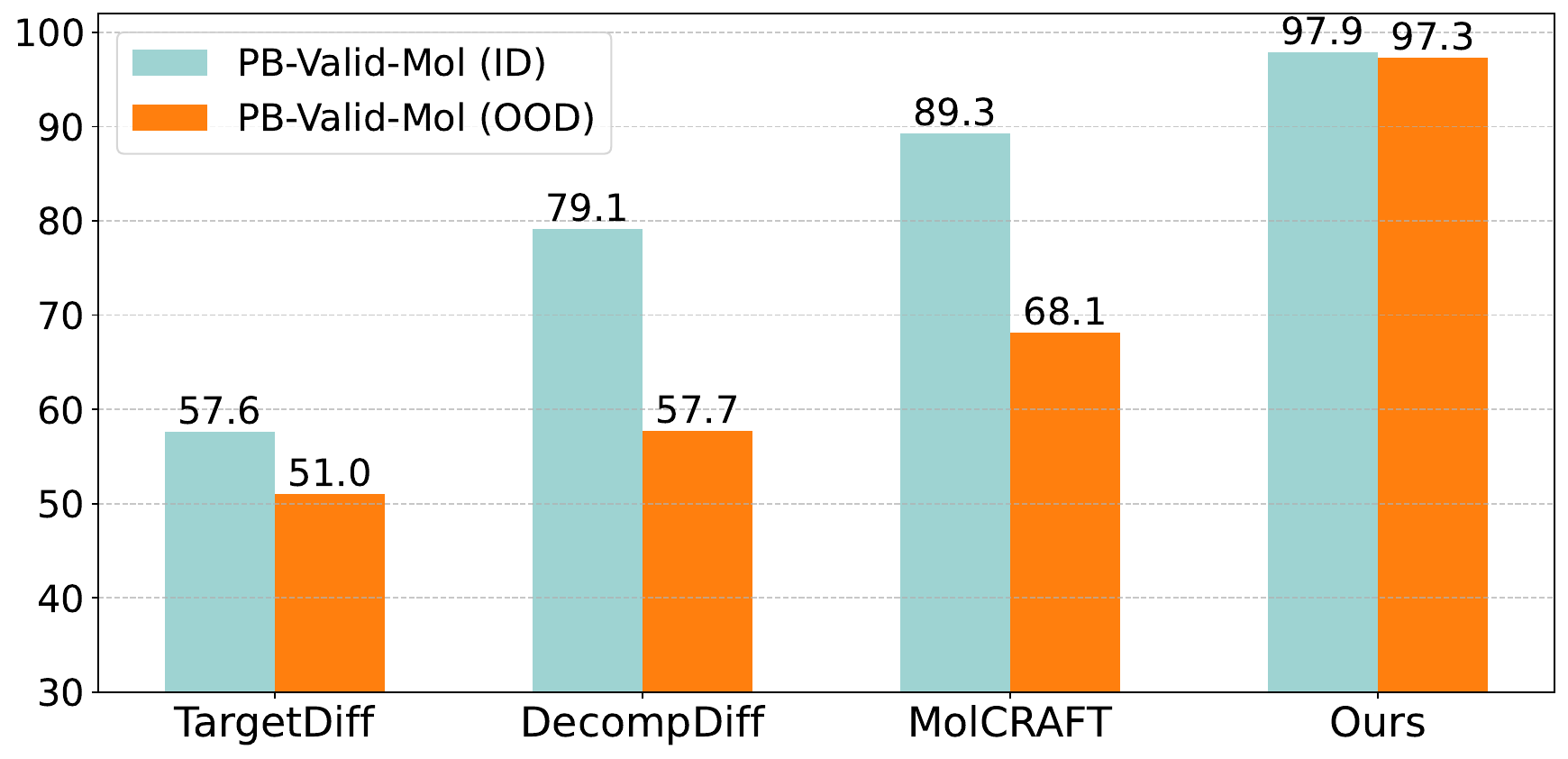}
    \vspace{-15pt}
    \caption{PoseBusters passing rates (\%) for non-autoregressive models, where ours maintains the best ID and OOD performance. \emph{ID:} in-distributional CrossDock test; \emph{OOD:} out-of-distributional PoseBusters test. \emph{PB-Valid-Mol:} intramolecular validity. Detailed results can be found in Appendix~\ref{sec:app-exp}.}
    \label{fig:pb_valid}
    \vspace{-15pt}
\end{figure}

Recent advances in geometric deep generative models have centered on non-autoregressive methods such as diffusion \citep{guan_decompdiff_2023} and Bayesian Flow Network (BFN) \citep{qu2024molcraft}, showing promise by capturing the structures at the global level. However, when evaluated with a rigorous out-of-distribution (OOD) test on PoseBusters benchmark \citep{buttenschoen2024posebusters}, Fig.~\ref{fig:pb_valid} shows a notable performance drop in intramolecular validity, suggesting that current global generation strategies may inadequately capture the fine-grained molecular geometries.

\begin{figure*}[t]
    \centering
    \includegraphics[width=\linewidth]{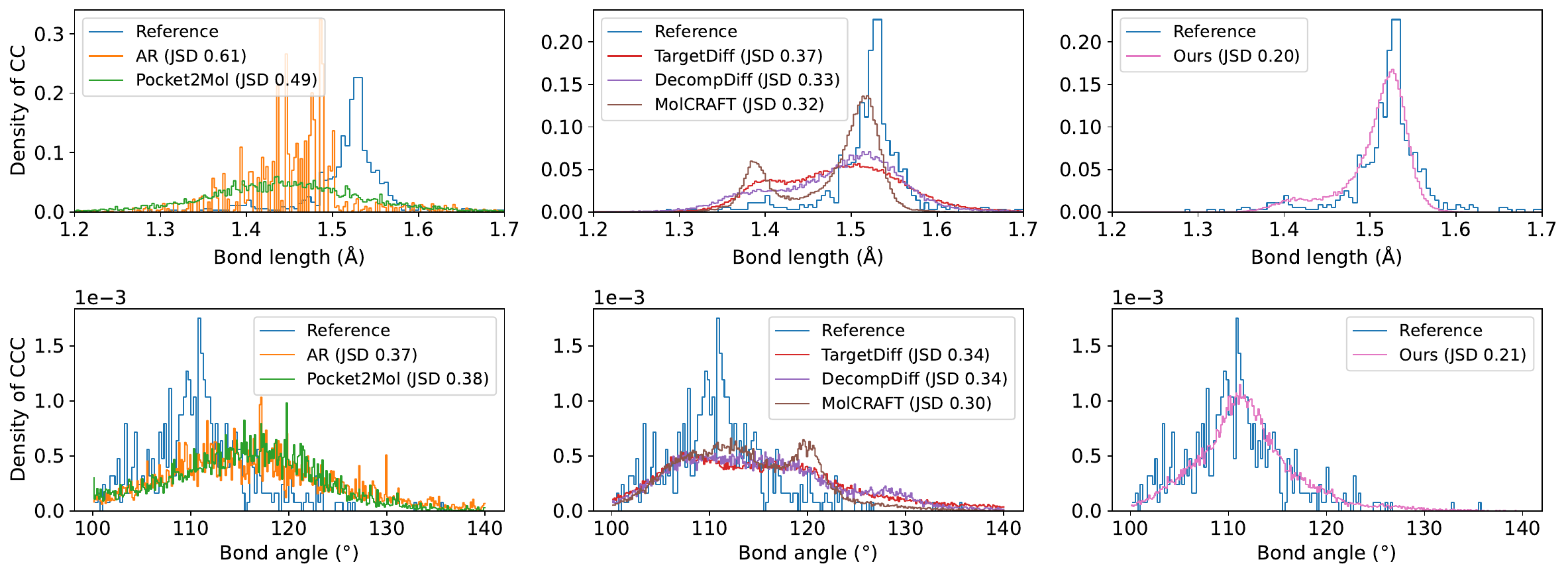}
    \vspace{-10pt}
    \caption{Visualization of the bond length and bond angle distribution for the most frequent bond types in CrossDock test. More types are shown in Appendix~\ref{subsec:app-exp-id}, where our \ours~consistently captures the molecular geometries most accurately.}
    \label{fig:top1_bond}
    \vspace{-5pt}
\end{figure*}

One key factor in geometric structure modeling is the intertwined probability path of different modalities: continuous atom positions and discrete molecular topologies. While these modalities jointly determine molecular geometries and protein-ligand interaction types, there lacks a systematic understanding in designing the twisted probability path for SBDD.
Prior works \citep{peng2023moldiff, vignac2023midimixedgraph3d, guan_decompdiff_2023} adopt sophisticated noise schedules adapted for different modalities, and propose to generate 3D positions first.
In our preliminary studies (Sec.~\ref{sec:preliminary}), we identify the potential problem of prioritizing the 3D modality, where the model cannot effectively utilize 2D topology information in the generation, suggesting that the current noise schedule is suboptimal. Despite its importance, the design of optimal noise schedules for twisted probability path remains a largely underexplored area.

To address the gap left by previous methods, we aim at a systematic solution for \emph{the optimal schedule for the twisted probability path of different modalities}, and emphasize the need for principled metrics relevant to generative modeling in order to evaluate schedule quality.
EquiFM \citep{song2024equivariant} introduces improved hybrid probability paths aligned with information-theoretic heuristics but lacks a theoretical foundation for optimality. To rigorously define optimality, we analyze the Variational Lower Bound (VLB) under varying noise schedules. Crucially, we prove that in the multi-modality generation, the VLB becomes a path-dependent integral; its value depends on the entire noise schedule, not just endpoints \citep{kingma2021variational}. 

This motivates our \textbf{VLB-Optimal Scheduling} (VOS), eliminating heuristic noise schedule design and directly linking it to the theoretical guarantees of the VLB landscape.
Specifically, we combat the \emph{combinatorial complexity in schedule design space} when navigating and optimizing for VLB.
To make it tractable to exhaust the possibilities, we demonstrate that the design space of multi-modality noise schedules can be reduced to a two-dimensional plane. Building on this insight, we develop a generalized objective, showing that decoupling timesteps during training allows a single model to implicitly cover this space, generalizing beyond fixed schedules. This framework enables efficient interpolation and extrapolation of schedules at inference time, bypassing costly retraining for new design constraints.
By advancing the underexplored area of modality-specific optimal scheduling, we address key shortcomings in SBDD models such as strained conformations and suboptimal interactions, substantially improving the geometric structure modeling. 

Our contributions can be summarized as follows:
\vspace{-10pt}
\begin{itemize}
    \item We introduce VLB-Optimal Scheduling (VOS), a novel method for systematic noise schedule design in SBDD, achieving fine-grained control over multi-modality interdependence and improved interaction modeling. 
\vspace{-5pt}
    \item We establish the theoretical link between noise schedules and VLB in multi-modality probabilistic modeling. Unlike prior heuristic approaches, our principled method reveals path-dependent VLB dynamics and provides insights for modality-specific scheduling.
    \item Integrated with advanced frameworks, our proposed \ours~achieves SOTA in de novo design with a remarkable PoseBusters passing rate of 95.9\% on CrossDock, and competitive in local docking with 44.0\% RMSD $<$ 2\r{A} on PoseBusters. 
\end{itemize}

\section{Issues with Current Probability Path}\label{sec:preliminary}
In this section, we introduce the formulation of Structure-Based Drug Design (SBDD) and demonstrate that the default noise schedule in current generative models leads to a 3D-dominant probability path. Through preliminary experiments, we demonstrate that the model trained along this path cannot adequately capture the interdependence between 2D and 3D modalities, which motivates the need for an optimal probability path facilitated by corresponding noise schedule.

\subsection{Problem Formulation}
Structure-based Drug Design (SBDD) involves modeling the conditional probability \(P(\rvx_M | \rvx_P)\), where \(\rvx_M = (\vr_M, \vh_M, \mA_M)\) represents the \(N\)-atom molecular geometry, and \(\rvx_P = (\vr_P, \vh_P)\) represents the protein target. Here, \(\vr \in \R^{N \times 3}\) denotes continuous atom positions, \(\vh \in \R^{N \times K_h}\) encodes discrete atom types, and \(\mA \in \R^{N \times (N-1) \times K_A}\) encodes discrete bond types, with \(K_h\) and \(K_A\) denoting the number of atom types and bond types, respectively. 
In molecular generation, $\vr$ from the continuous modality is usually described as the 3D geometry, and $(\vh, \mA)$ from the discrete modality as the 2D topology of the molecular graph.

\begin{figure}[htbp]
    \centering
    \includegraphics[width=\linewidth]{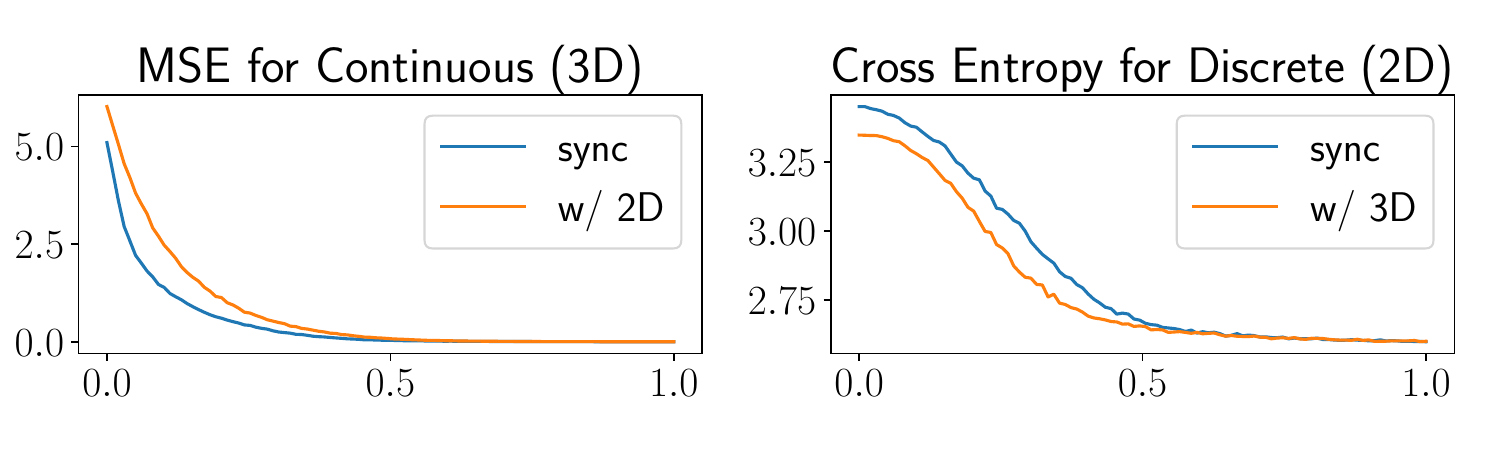}
    \vspace{-20pt}
    \caption{Validation loss curves on default schedule w.r.t. time in generation. \emph{Sync:} Modalities at the same timestep. \emph{w/ 2D:} Discrete modality at $t=1$. \emph{w/ 3D:} Continuous modality at $t=1$.}
    \label{fig:loss_preliminary}
\end{figure}

\subsection{3D-driven Probability Path}

Fig.~\ref{fig:top1_bond} suggests that models have yet to accurately capture the molecular geometry, potentially resulting in 3D molecular structures that are incompatible with their 2D molecular graphs (Appendix~\ref{sec:app-exp}). To investigate this, we visualize the modality-specific validation losses for a vanilla BFN \citep{qu2024molcraft} trained with default noise schedule.

As shown in Fig.~\ref{fig:loss_preliminary}, the continuous modality loss decreases before the discrete modality, verifying that the twisted probability path is driven by 3D modality.
However, the problem lies in that the model leverages cleaner 3D input to denoise the 2D topology effectively (\emph{Right}), but fails the other way around (\emph{Left}), as it performs worse when exposed to less noisy 2D input, suggesting that it is unable to benefit from cleaner 2D information. 
This highlights that the current noise schedule favors a twisted probability path driven by 3D modality, inducing a generative process functioning with lower-noise 3D conformation and higher-noise 2D topology.

Intuitively, a well-balanced probability path should allow the models to leverage cleaner input from either modality to inform the generative process effectively, thereby accurately capturing the molecular geometries required for effective drug design.
The problem of current twisted probability path is strengthened by the observation of inaccurate geometric structure modeling, where SBDD baselines exhibit abnormal bond lengths and angles (Fig.~\ref{fig:other_len}-\ref{fig:other_torsion}), suggesting inconsistency between modalities. 
We hypothesize that the distorted molecular geometry originates from the gap between modalities in the twisted probability path, which inspires our search for an optimal scheduling.

We desire the optimal probability path defined by the optimal noise schedule to benefit from mutually informed modalities,
with the potential to improve the generation of chemically valid and spatially accurate molecules, ultimately enhancing protein-ligand interaction modeling.

\begin{figure*}[t]
    \centering
    \includegraphics[width=\linewidth]{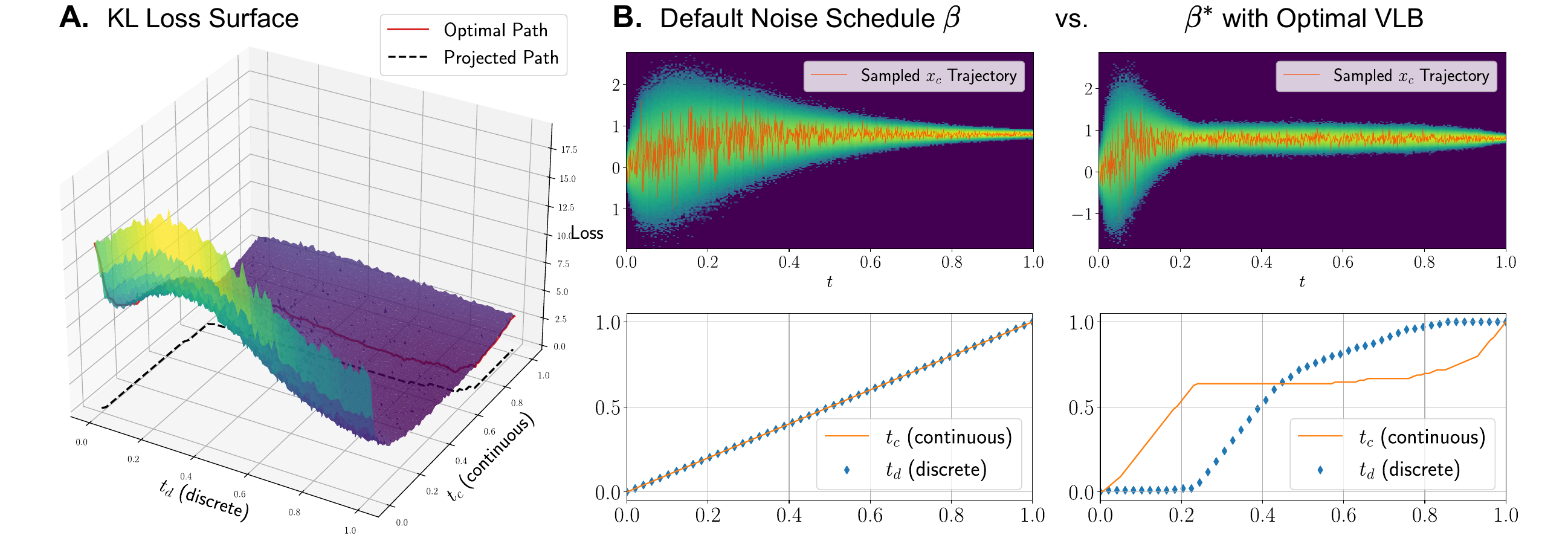}
    \caption{Our proposed VLB-Optimal Scheduling (VOS) that works by estimating the loss landscape and deriving an optimal noise schedule. \textbf{A.} Visualization of the loss surface over the function space. \textbf{B.} \emph{Upper:} Visualization of the probability path of continuous data $\rvx_c$. \emph{Lower:} The equivalent time-rescaling functions $t_c\equiv f(t), t_d\equiv g(t)$.}
    \label{fig:surface}
    \vspace{-10pt}
\end{figure*}

\section{Background}
Our work is built within the framework of Bayesian Flow Network (BFN) \citep{graves2023bayesian}, the SOTA model for 3D molecular generation \citep{song2024unified, qu2024molcraft} that shows superiority over diffusion counterparts. In this section, we introduce the noise schedule and define the corresponding noising process, paving the way for Variational Lower Bound (VLB) analysis w.r.t. noise schedules.

Similar to diffusion, BFN involves a noising process $q$ for data $\rvx$ to obtain a temporal sequence of observed latents $\rvy_{1:n} := \{\rvy_1, \dots, \rvy_n \}$, and optimizes the VLB:
\begin{align}
    \log p(\rvx) &\ge \text{VLB} 
    = \nonumber \E_{q(\rvy_{1:n}\mid\rvx)}[\log p(\rvx\mid\rvy_{1:n}) \\ & 
    - \KL(q(\rvy_{1:n}|\rvx) \Vert p(\rvy_{1:n}))]
\end{align}
where the noising process $q$ is defined by the variational distribution $q(\rvy_{1:n}\mid\rvx)=\prod_{i=1}^n q(\rvy_i \mid \rvx)$ that depends on the noise schedule over timesteps:
\begin{equation}\label{eq:q}
    q(\rvy_i \mid \rvx) = \begin{cases}
        \gN(\rvx, \alpha_i^{-1}\mI), &\text{continuous data} \\
        \gN(\alpha_i(K{\rve_\rvx} - \mathbf{1}), \alpha_i K\mI), &\text{discrete one-hot}
    \end{cases}
\end{equation}
where $K$ is the number of classes, $\rve_{\rvx}$ is Kronecker function, i.e., the projection from a class index $\mathbf{x}=j$ to a one-hot vector $\in \mathbb{R}^K$ with the $j$-th value equal to 1. For time step $i$, the discretized $\alpha_i$ is obtained from modality-specific noise schedules, defined as monotonically increasing differentiable functions over $t \in [0, 1]$:
\begin{equation}\label{eq:beta}
\beta_{c}(t) = \sigma_1^{-2t} - 1, \quad \beta_{d}(t) = \beta_{1} t^2,
\end{equation}
where \(\sigma_1 \in \R^+\) controls the noise for continuous atomic positions, and \(\beta_1 \in \R^+\) is the hyperparameter for discrete topology. The noise level \(\alpha(t)=\beta'(t):=d\beta(t) / dt\).

BFN differs from diffusions in the generative process, where instead of noisy latents $\rvy$, the network is informed by a lower-variance posterior $\vtheta$ given the observed latents, from the Bayesian flow distribution of Gaussian or Dirac $\delta$:
\begin{equation}
    p_F(\vtheta \mid \rvx; t) = \begin{cases}
        \gN(\gamma\rvx, \gamma(1-\gamma)\mI), &\text{continuous} \\
        \delta(\vtheta - \text{softmax}(\rvy)), &\text{discrete}
    \end{cases}
\end{equation}
where $\gamma := \frac{\beta_c(t)}{1+\beta_c(t)}$, $\rvy$ follows the discrete case in Eq.~\ref{eq:q}.

The network $\tilde{\rvx}_\phi(\vtheta, t)$\footnote{The actual noise prediction model is specified as $\tilde{\rvx}_\phi(\vtheta, t, \rvx_P)$ with the protein structure input. We omit $\rvx_P$ for simplicity.} is trained to denoise $\rvx$ given $\vtheta$ instead of $\rvy$, and thus constructs a receiver distribution $p_R(\rvy \mid \vtheta; t)=q(\rvy \mid\tilde{\rvx}_\phi(\vtheta, t))$. 
We show in Appendix~\ref{subsec:app-vlb} that the single-modality VLB objective simplifies to:
\begin{align}\label{eq:vlb_bfn}
    \gL_{\text{VLB}}(\rvx)&=-\E_{q(\rvy_{1:n}\mid\rvx)}\log p(\rvx\mid\rvy_{1:n}) \nonumber \\
    +\sum_{i=1}^n&\KL(q(\rvy_{i}\mid\rvx) ~\Vert~ p(\rvy_{i}\mid \rvy_{1:i-1}))
\end{align}
and can be optimized by the continuous-time loss:
\begin{align}\label{eq:kl_bfn}
\gL^\infty(\rvx) 
&= \underset{t\sim U(0,1), p_F(\vtheta | \rvx; t)}{\E} \KL(q(\rvy~|~\rvx; t)~\Vert~p_R(\rvy~|~\vtheta; t)) \nonumber \\
&= \frac{1}{2}\underset{t\sim U(0,1), p_F(\vtheta | \rvx; t)}{\E} \boldsymbol{\beta}'(t)\Vert \rvx - \tilde{\rvx}_\phi(\vtheta, t)\Vert^2
\end{align}
where we redefine $\rvx:=\sqrt{K}\rve_\rvx$ for the discrete data for simplicity. 
Based on the monotonic and thus invertible function $u\equiv\beta(t)$, we perform a change of variables similarly to \citet{kingma2021variational} as $t=\beta^{-1}(u)$, and rewrite the network as $\tilde{\rvx}_\phi(\vtheta, u)$. Therefore, the loss in Eq.~\ref{eq:kl_bfn} is equivalent to the continuous-time loss expressed by noise schedule $\beta(t): [0,1] \to \R^+$:
\begin{align}\label{eq:kl_beta}
\gL^\infty(\rvx) =  \frac{1}{2}\int_{\beta(0)}^{\beta(1)}\underset{p_F(\vtheta | \rvx; u)}{\E}\Vert \rvx - \tilde{\rvx}_\phi(\vtheta, u)\Vert^2 du.
\end{align}

In the single-modality case, this states that the VLB is invariant to the shape of noise schedule functions except for the endpoints \citep{kingma2021variational}. However, we will see in the following section that this invariance no longer holds for multi-modality cases, where the VLB becomes:
\begin{align}\label{eq:vlb_joint_modality}
    \gL_{\text{VLB}}(\rvx) &= \frac{1}{2}\mathbb{E}_{p_F(\vtheta \mid \rvx; t)} \int_0^1 [\beta_c'(t) \Vert \rvx_c - \tilde{\rvx}_{\phi,c}(\vtheta, \boldsymbol{\beta}(t)) \Vert^2 \nonumber \\
    &+ \beta_d'(t) \Vert \rvx_d - \tilde{\rvx}_{\phi,d}(\vtheta, \boldsymbol{\beta}(t)) \Vert^2 ] dt,
\end{align}
which is simply summed along each modality due to the factorized nature of $q$, yet this results in path-dependent VLB which is closely related to the design of joint noise schedule $\boldsymbol{\beta}(t): [0,1] \to (\R^+)^2$.


\section{Methodology}
We propose VLB-Optimal Scheduling (VOS), a principled methodology for optimal probability path in SBDD, by analyzing and identifying the optimal noise schedule for both discrete 2D topologies and continuous 3D atomic positions. 

\paragraph{Overview}
(1) In Sec.~\ref{subsec:vlb_wrt_schedule}, we theoretically analyze the path-dependent VLB for a given twisted probability path induced by specific noise schedule in that function space.
(2) In Sec.~\ref{subsec:comb_space}, we formalize the entire function space of possible noise schedules, and propose to navigate the space by time rescaling.
(3) In Sec.~\ref{subsec:dp_vos}, we propose a path-invariant generalized loss objective that enables efficient estimation of the VLB landscape over the entire function space of noise schedules. Then, we describe the method to search for the optimal schedule on the landscape that maximizes the VLB.

\subsection{Path-dependent VLB for Joint Noise Schedule}\label{subsec:vlb_wrt_schedule}
In this section, we establish a key result for the twisted probability path, where the joint noise schedule $\boldsymbol{\beta}(t): [0,1] \to (\R^+)^2$ induces path-dependent VLB.

A key foundation in single modality is that the continuous-time loss, i.e., VLB, remains invariant to the shape of noise schedule function $\beta(t)$, except for the endpoints $\beta(0), \beta(1)$ \citep{kingma2021variational}, allowing the design of different schedules for efficient training. As a natural extension, we generalize this invariance to multi-modalities given $\beta_c\ne\beta_d$:
\begin{equation}\label{eq:generalized_loss}
    \dot{\gL}^\infty(\rvx) = \frac{1}{2} \int_{\beta_c(0)}^{\beta_c(1)}\int_{\beta_d(0)}^{\beta_d(1)}\underset{p_F(\vtheta | \rvx; \boldsymbol{\beta})}{\E}\Vert \rvx - \tilde{\rvx}_\phi(\vtheta, \boldsymbol{\beta})\Vert^2 d\boldsymbol{\beta}.
\end{equation}
This equation shows that the generalized loss is invariant to the decoupled schedules as a surface integral over the plane $[\beta_c(0), \beta_c(1)] \times [\beta_d(0), \beta_d(1)]$.

However, this generalized loss $\dot{L}$ no longer corresponds to the objective of generative modeling, where the VLB should be a path integral along a line on such plane:
\begin{equation}\label{eq:VLB}
    \gL^\infty(\rvx) =  \frac{1}{2}\int_{\beta_c, \beta_d}\underset{p_F(\vtheta | \rvx; \boldsymbol{\beta})}{\E}\Vert \rvx - \tilde{\rvx}_\phi(\vtheta, \boldsymbol{\beta})\Vert^2 d\boldsymbol{\beta}.
\end{equation}
This curve is equivalent to a specific coupled noise schedule $\boldsymbol{\beta} \in \gZ$. Therefore, we conclude that the VLB varies for different coupling even with the same endpoints $\boldsymbol{\beta}(0)=(\beta_c(0), \beta_d(0))$ and $\boldsymbol{\beta}(1)=(\beta_c(1), \beta_d(1))$, and it is no longer agnostic to the intermediate trajectory of $\boldsymbol{\beta}(t)$, which is essentially different from the generative modeling within single modality.

The fundamental challenge, then, lies in identifying an optimal joint schedule $\boldsymbol{\beta}^*$ in the design space $\gZ$ with the best possible VLB integrated along that path:
\begin{equation}
    \boldsymbol{\beta}^*=\argmin_{\boldsymbol{\beta} \in \gZ}  \int_{\boldsymbol{\beta}}\underset{p_F(\vtheta | \rvx; \boldsymbol{\beta})}{\E}\Vert \rvx - \tilde{\rvx}_\phi(\vtheta, \boldsymbol{\beta})\Vert^2 d\boldsymbol{\beta}.
\end{equation}


\begin{algorithm}[t]
\caption{Deriving Optimal Schedule}
\label{algo:optimal_schedule}
\KwIn{Multi-modality data $\rvx$, default noise schedule $\boldsymbol{\beta}=(\tilde{\beta}_c$, $\tilde{\beta}_d)$, grid resolution $M$, step $N$, step scale $K$.}
\KwOut{Optimal schedule $\boldsymbol{\beta}^*$, generative model $\tilde{\rvx}_\phi(\vtheta, \boldsymbol{t})$.}


Train $\tilde{\rvx}_\phi(\vtheta,t)$ to minimize the generalized loss $\dot{\gL}$ (Eq.~\ref{eq:generalized_loss_for_t})\;
\(
\dot{\gL}^\infty(\rvx) = \frac{1}{2}\int_{0}^{1}\int_{0}^{1} \underset{p_F(\vtheta | \rvx; t)}{\E}\Vert \rvx - \tilde{\rvx}_\phi(\vtheta, \boldsymbol{t})\Vert^2 dt_c dt_d
\)


Discretized grid $g=(t_c, t_d)_{M\times M} \gets [\frac{1}{M}, \dots, 1]^2$\;

\ForEach{$(t_c, t_d)$ in $g$}{
    Compute cost $C(t_c, t_d) \gets \frac{1}{2} \Vert \rvx - \tilde{\rvx}_\phi(\vtheta, t) \Vert^2$
}

Optimal path $\{t_c^*\equiv f(t), t_d^*\equiv g(t)\}, J^* \gets \mathrm{DP}(C, N, K)$\;

Optimal schedule $\boldsymbol\beta^*(t) \gets (\tilde{\beta}_c(t_c^*), \tilde{\beta}_d(t_d^*))$\;
\end{algorithm}

\subsection{Design Space of Joint Noise Schedules}\label{subsec:comb_space}

To formalize the design space of noise schedules, we define the space of monotonically increasing functions $\gX:[0,1]\to \R^+$, thus the coupled function space $\gZ$ is:
\begin{equation}
    \{ \boldsymbol{\beta}(t)= (\beta_{c}(t), \beta_{d}(t)) \mid \forall \beta_c, \beta_d \in \gX, \beta_c' >0, \beta_d'>0 \},
\end{equation}
where \(\beta_{c}(t)\) and \(\beta_{d}(t)\) are monotonic schedules for the continuous and discrete modalities, respectively. 

To explore arbitrary schedule configurations, we introduce time-rescaling functions $t_{c}\equiv f(t)$, $t_{d}\equiv g(t)$, allowing for arbitrary $\boldsymbol{\beta}$ expressed in predefined $\tilde{\beta}_{c}, \tilde{\beta}_{d}$ as in Eq.~\ref{eq:beta} and varying forms of implicit functions $f$ and $g$ by:
\begin{equation}\label{eq:space}
    \boldsymbol{\beta}(t) = (\tilde{\beta}_c(f(t)), \tilde{\beta}_{d}(g(t)))
\end{equation}
for which we have the following theorem that guarantees this general form of $\boldsymbol{\beta}$ expressed in terms of $t_c, t_d$ is sufficient to cover the entire function space $\gZ$.
\begin{theorem}
Suppose we have a monotonic function $\tilde{\beta}_{m}(t): [0,1]\to \R^+$, and let $\beta_{m}(t)$ be any such monotonic function. Then there exists a time-rescaling function 
$t_m \equiv f(t)$ such that $\beta_{m}(t)=\tilde{\beta}_{m}(t_m)$.
In fact, $f(t)=\tilde{\beta}^{-1}_{m}(\beta_{m}(t))$ has the same monotonicity as $\beta_m(t), \tilde\beta_m(t)$.
\end{theorem}

\begin{remark}\label{rem:time_decouple}
It follows from the monotonicity of the schedule functions that we can obtain arbitrary combination of noise levels simply by setting $t_{c}, t_{d} \in [0,1]$ separately.
\end{remark}


\subsection{Navigating for the VLB-optimal Schedule}\label{subsec:dp_vos}
The first obstacle in identifying the optimal $\boldsymbol{\beta}^*$ is to obtain a model $\tilde{\rvx}_\phi(\vtheta, \boldsymbol{\beta})$ that can be evaluated over all possible $\boldsymbol{\beta} \in \gZ$.
We facilitate the VLB analysis of different joint noise schedules by training $\tilde{\rvx}_\phi(\vtheta, \boldsymbol{t})$ to minimize the generalized loss $\dot{\gL}$ described in Eq.~\ref{eq:generalized_loss} through the change of variables $t_c\equiv \beta_c^{-1}(t)$, $t_d\equiv \beta_d^{-1}(t)$, thereby the objective becomes:
\begin{align}\label{eq:generalized_loss_for_t}
\dot{\gL}^\infty(\rvx) = \frac{1}{2}\int_{0}^{1}\int_{0}^{1} \underset{p_F(\vtheta | \rvx; \boldsymbol{t})}{\E}\Vert \rvx - \tilde{\rvx}_\phi(\vtheta, \boldsymbol{t})\Vert^2 dt_c dt_d,
\end{align}
for which the proposition holds (proof in Appendix~\ref{subsec:app-vlb-line-integral}):
\begin{proposition}\label{prop:vlb}
Suppose we have a model $\tilde{\rvx}_\phi(\vtheta, \boldsymbol{t})$ trained by Eq.~\ref{eq:generalized_loss_for_t}, and let $\beta_c(t), \beta_d(t)$ be any monotonically increasing functions in $\gX$. Then the line integral in Eq.~\ref{eq:VLB} corresponds to the negative VLB for $\boldsymbol{\beta}(t)=(\beta_c(t), \beta_d(t))$.
\end{proposition}

Then, we present the method to navigate the function space of arbitrary $\boldsymbol{\beta}$.
Note that from Remark~\ref{rem:time_decouple}, we can discretize the function space through discretized combinations of $t_c, t_d$.
Therefore, we can formulate finding the optimal $\boldsymbol{\beta}^*$ as a search problem for the solution with the minimal cumulative cost along the discretized $N$-step trajectory \( \{t_c, t_d\} \) from \( [0, 0]\) to \([1, 1]\).

Depicted in Fig.~\ref{fig:surface}, we estimate the cost matrix across a grid of possible \( t_c \) and \( t_d \) values by evaluating the KL divergence over a batch of samples given the $\dot{\gL}$-trained model $\tilde{\rvx}_\phi(\vtheta, \boldsymbol{\beta})$, and fit a smooth loss surface using B-spline interpolation.

\begin{definition}[Cost]
    The cost at a given noise level \(t_c, t_d\) as implicit functions over $t$ is defined as:
    \begin{align}
    C(t_c, t_d) 
    &= \frac{1}{2} \Vert \rvx - \tilde{\rvx}_\phi(\vtheta, \boldsymbol{\beta}) \Vert^2
    \end{align}
\end{definition}

We employ dynamic programming to solve the search problem for a minimal cumulative cost $J$:
\begin{equation}
    J(t_c, t_d) = \min_{(\eps_c, \eps_d)} \big(J(t_c-\eps_c, t_d-\eps_d) + \boldsymbol{\alpha} C(t_c, t_d)\big).
\end{equation}
where $\boldsymbol{\alpha}:=\boldsymbol{\beta}'(t)$ is the accuracy level,
and the valid ranges of $(\eps_c, \eps_d)$ are approximated by the gradient of the smooth surface.
It is guaranteed that the method yields a schedule with maximal VLB given the corresponding accuracy level:
\begin{remark}
    $J(1, 1)$ corresponds to an unbiased Monte-Carlo estimate of the optimal VLB among all generative models trained by arbitrary $\boldsymbol{\beta} \in \gZ$, and backtracking the path $\{t_c=f(t), t_d=g(t)\}$ from $[1, 1]$ to $[0, 0]$ yields the optimal schedule $\boldsymbol{\beta}^*(t) = (\tilde{\beta}_c(t_c), \tilde{\beta}_d(t_d))$.
\end{remark}

Table~\ref{tab:VLB_comparison} empirically validates the derived schedule with better VLB than default. The overall procedure is summarized in Algorithm~\ref{algo:optimal_schedule}, with function $\mathrm{DP}$ specified in Algorithm~\ref{algo:dp}.

We visualize the derived optimal schedule in Fig.~\ref{fig:surface}B. Intuitively, this time rescaling corresponds to a two-stage probability path: (1) \emph{Shape-driven sketching:} First, the generation predominantly focuses on generating the continuous atom positions and largely ignores the discrete topology. Then, when $t \in [0.3, 0.8]$, the model starts to fit a possible 2D molecular graph into the rough shape of fixed 3D conformation. (2) \emph{Topology-driven docking:} at the last stage $t > 0.8$, the generation enters into the docking stage, altering the conformation according to the discrete molecular topologies. By facilitating such a principled probability path, the optimal noise schedule effectively harnesses different modalities, achieving the best VLB and sample quality.

\section{Experiments}\label{sec:exp}
\subsection{Experimental Setup}\label{subsec:exp-setup}
We conduct two main experiments within the broader scope of SBDD: (1) de novo design in the in-distributional (ID) and out-of-distributional (OOD) settings, and (2) molecular docking.
The same model checkpoint trained with generalized loss is evaluated throughout both main experiments.

\paragraph{Dataset.} Following SBDD conventions \citep{luo_3d_2022}, we adopt the same split of CrossDock \citep{francoeur2020three} to train and validate our model, which consists of 100,000 training poses and 100 validation poses. 
(1) For de novo design, we evaluate on an OOD subset of PoseBusters \citep{buttenschoen2024posebusters} in addition to the ID CrossDock test set. We cluster all chains via MMseqs2 \citep{steinegger2017mmseqs2} and filter out test proteins with any chain $>$ 30\% sequence identity w.r.t. CrossDock training sequences, obtaining 180 test proteins.
This avoids possible information leakage from the original data splits (detailed discussion in Appendix~\ref{subsec:app-problem-cd}), and serves as a more reliable held-out test, where the structures of protein-ligand complexes are experimentally determined.
(2) For molecular docking, after removing 10 proteins that cannot be processed for non-standard residues from the PoseBusters V2, we test on the remaining 298 protein-ligand complexes as ground-truth-based evaluation.

\paragraph{Baselines.} We consider the following SBDD baselines: 
(1) Autoregressive models including AR \citep{luo_3d_2022}, Pocket2Mol \citep{peng_pocket2mol_2022}, 
(2) Diffusion-based TargetDiff \citep{guan_3d_2023}, DecompDiff \citep{guan_decompdiff_2023}, and
(3) BFN-based MolCRAFT \citep{qu2024molcraft}.
A detailed list of baselines can be referred to in Appendix~\ref{subsec:app-exp-setup}.

\begin{table*}[t]
\caption{Performance on CrossDock in an in-distribution (ID) setting and PoseBusters in an out-of-distribution (OOD) setting, where \ours~shows robust results.
$\heartsuit$: results cited from \citet{qu2024molcraft}.
$\dagger$: results calculated by us using the official samples.
$\diamondsuit$: results calculated by us using the official code. Top-2 results highlighted in \textbf{bold} and \ul{underlined}, respectively. }
\centering
\label{tab:main}
\resizebox{\linewidth}{!}{%
\begin{tabular}{l|c|cc|cc|cc|c|c|c|c|c|c|c}
\toprule
\multirow{2}{*}{Methods} & PB-Valid$\dagger$ & \multicolumn{2}{c|}{Vina Score ($\downarrow$)} & \multicolumn{2}{c|}{Vina Min ($\downarrow$)} & \multicolumn{2}{c|}{Vina Dock ($\downarrow$)} & scRMSD & Energy                      & Connected$\dagger$ & QED & SA & Div & Size \\
                         & Avg. ($\uparrow$)                 & Avg.                   & Med.                  & Avg.                  & Med.                 & Avg.                  & Med.                  & \textless 2 \r{A} ($\uparrow$) & Passed$\dagger$ ($\uparrow$) & Avg. ($\uparrow$)              & Avg.    & Avg.        & Avg.                    & Avg. \\ \midrule
CrossDock (ID)$\heartsuit$                & 95.0\%                & -6.36                  & -6.46                 & -6.71                 & -6.49                & -7.45                 & -7.26                 & 34.0\% & 98.0\%                                  & -                & 0.48             & 0.73  & -    & 22.8 \\ \midrule
AR$\heartsuit$                       & 59.0\%                & -5.75                  & -5.64                 & -6.18                 & -5.88                & -6.75                 & -6.62                 & 36.5\% & 84.9\%                                & 93.5\%              & 0.51             & 0.63    & 0.70    & 17.7 \\
Pocket2Mol$\heartsuit$               & 72.3\%                & -5.14                  & -4.70                 & -6.42                 & -5.82                & -7.15                 & -6.79                 & 32.0\% & \ul{97.3\%}                      & 96.3\%                & {0.57}       & 0.76 & 0.69     & 17.7 \\
FLAG$\heartsuit$                     & 16.0\%                & 45.85                  & 36.52                 & 9.71                  & -2.43                & -4.84                 & -5.56                 & 0.3\% & 83.4\%                                & 97.1\%         & {0.61}    & 0.63   & 0.70           & 16.7 \\
TargetDiff$\heartsuit$               & 50.5\%                & -5.47                  & -6.30                 & -6.64                 & -6.83                & -7.80                 & -7.91                 & 37.1\% & 69.8\%                                   & 90.4\%      & 0.48             & 0.58   & 0.72    & 24.2 \\
DiffSBDD$\dagger$                 & 37.6\%                & -1.44                  & -4.91                 & -4.52                 & -5.84                & -7.14                 & -7.3                  & 18.7\% & 74.0\%                             & 93.2\%   & 0.47             & 0.58   & 0.73          & 24.4 \\
DecompDiff$\heartsuit$               & 71.7\%                & -5.19                  & -5.27                 & -6.03                 & -6.00                & -7.03                 & -7.16                 & 24.2\% & 84.1\%                                & 82.9\%      & 0.51             & 0.66  & 0.73     & 21.2 \\
MolCRAFT$\heartsuit$                 & \ul{84.6\%}          & \ul{-6.55}            & \ul{-6.95}           & \ul{-7.21}           & \ul{-7.14}          & \ul{-7.67}           & \ul{-7.82}           & \textbf{46.8\%} & 91.1\%                                   & \ul{96.7\%}     & 0.50             & 0.67  & 0.72      & 22.7 \\
\rowcolor{gray!20}
\ours~(Ours)                     & \textbf{95.9\%}       & \textbf{-6.88}         & \textbf{-7.03}        & \textbf{-7.23}        & \textbf{-7.27}       &  \textbf{-7.92}                     &  \textbf{-7.92}                     &     \ul{41.1\%} & \textbf{98.5\%}                       & \textbf{97.9\%}                & 0.56             & {0.74}  & 0.69   & 22.6 \\ \midrule\midrule
PoseBusters (OOD)$\dagger$                & 98.9\%                & -7.06                  & -7.05                 & -7.50                 & -7.41                & -7.98                 & -7.82                 & 59.4\%                                                                     & 100\%           & -         & 0.40 & 0.72 & - & 25.7 \\ \midrule
AR$\diamondsuit$                       & 54.7\%                & -5.45                  & -5.17                 & -5.67                 & -5.38                & -6.18                 & -5.94                 & 35.3\%                                                                     & 77.7\%          & 39.1\%    & 0.50 & 0.67 & 0.76 & 13.6 \\
Pocket2Mol$\diamondsuit$               & \ul{63.6\%}          & -5.39                  & -5.03                 & -6.64                 & -6.24                & -7.40                 & -7.03                 & 37.8\%                                                                       & 97.4\%          & 67.7\%    & 0.57 & 0.74 & 0.73 & 17.4 \\
TargetDiff$\diamondsuit$               & 32.3\%                & -6.57                  & -6.78                 & -7.16                 & -7.31                & \ul{-8.18}           & \textbf{-8.20}        & 32.3\%                                                                     & 65.2\%          & 81.3\%    & 0.41 & 0.55 & 0.67 & 27.0 \\
DecompDiff$\diamondsuit$               & 40.2\%                & -3.14                  & -3.02                 & -4.03                 & -4.11                & -5.06                 & -5.40                 & 17.0\%                                                                     & 80.1\%          & 82.9\%    & 0.47 & 0.66 & 0.81 & 19.3 \\
MolCRAFT$\diamondsuit$                 & 57.8\%                & \ul{-7.29}            & \ul{-7.11}           & \ul{-7.44}           & \ul{-7.22}          & -7.95                 & -7.73                 & \ul{46.4\%}                                                               & 71.6\%          & \ul{97.0\%}    & 0.41 & 0.65 & 0.65 & 23.8 \\
\rowcolor{gray!20}
\ours~(Ours)                     & \textbf{79.1\%}       & \textbf{-7.59}         & \textbf{-7.54}        & \textbf{-7.74}        & \textbf{-7.67}       & \textbf{-8.20}        & \ul{-7.99}           & \textbf{56.1\%}                                                            & \textbf{98.1\%} & \textbf{97.3\%}    & 0.49 & 0.72 & 0.67 & 23.5 \\ \bottomrule
\end{tabular}
}
\end{table*}

\paragraph{Metrics.} We employ the following metrics: 
(1) For conformation quality, we report PoseBusters passing rate (\textbf{PB-Valid}), 
and \textbf{Strain Energy Passed} denotes the ratio of molecules passing PoseBusters internal energy check.
(2) For interaction modeling, we emphasize \textbf{RMSD}, the root-mean-squared distance between generated and ground truth pose in docking. We also report the self-consistency RMSD (\textbf{scRMSD}) in de novo design calculated between generated and Vina Dock pose as an indicator of binding mode consistency upon redocking.
Affinity metrics are calculated by AutoDock Vina \citep{eberhardt2021autodock}, including binding affinities where \textbf{Vina Score} directly scores the generated molecular pose in the pocket, \textbf{Vina Min} quickly optimizes the pose in-place and scores the minimized pose, and \textbf{Vina Dock} exhaustively searches for optimal pose to obtain the lowest energy. 
(3) For molecular properties, we report drug-likeness (\textbf{QED}) and synthetic accessibility (\textbf{SA}), which are desired to fall within reasonable ranges.
We additionally report the percentage of fully connected molecules (\textbf{Connected}) to show the performance of generation successes, together with the average number of atoms of those successfully generated molecules (\textbf{Size}).

\begin{figure*}
    \begin{minipage}{0.75\linewidth}
    \includegraphics[width=0.9\linewidth]{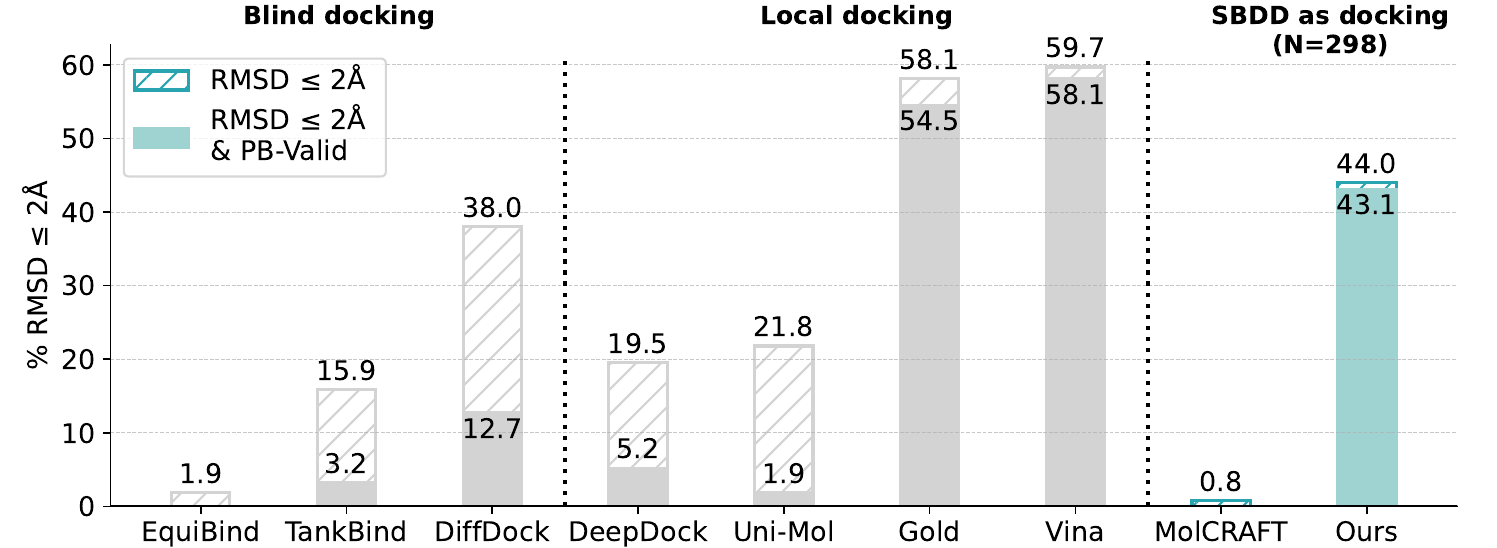}
    \caption{PoseBusters V2 structural accuracy (RMSD) and validity (PB-Valid) for docking methods (in {\color{gray}gray}, cited from \citet{abramson2024accurate}) and SBDD methods (in {\color{teal}blue}, calculated by us). }
    \label{fig:main-dock}
    \end{minipage}
\begin{minipage}{0.23\linewidth}
\centering
\vspace{-0.5cm}
\captionof{table}{Tanimoto similarity of interaction profiles between PoseBusters reference and generated PB-Valid molecules.}
\label{tab:interaction_sim}
\vskip 0.1in
\resizebox{0.85\linewidth}{!}{%
\begin{tabular}{l|c}
\toprule
Methods    & Sim. ($\uparrow$)           \\ \midrule
AR         & 0.221          \\
Pocket2Mol & 0.436          \\
TargetDiff & 0.458          \\
DecompDiff & 0.374          \\
MolCRAFT   & 0.498          \\
\rowcolor{gray!20}
\ours~(Ours)       & \textbf{0.551} \\ \bottomrule
\end{tabular}%
} 
\end{minipage}
\vspace{-10pt}
\end{figure*}

\subsection{De novo Design}

We report the main results for ID and OOD design in Table~\ref{tab:main} by sampling 100 molecules for each test protein.

\paragraph{\ours~achieves the most accurate molecular geometries.} For intramolecular validity, we achieve the highest passing rate of internal energy, demonstrating our conformation stability and showing robustness in both ID and OOD settings. We additionally report the bond length, angle, and torsion angle distributions in Fig.~\ref{fig:other_len}, \ref{fig:other_angle}, \ref{fig:other_torsion}, further underscoring its superiority. 

\paragraph{\ours~generates the best binding poses.}
For protein-ligand intermolecular validity, we excel at the highest PB-Valid, ensuring reasonable binding poses. For binding affinities, our method outperforms all other models with an average Vina Score of -6.88 (ID) and -7.45 kcal/mol (OOD), very close to Vina Min and Dock. 

\paragraph{CrossDock benchmark may not be challenging enough.}
In Table~\ref{tab:main}, we observe that our rate of PB-Valid and internal energy passed even outperforms the CrossDock test set, suggesting potential problems with this synthetic dataset \citep{qu2024molcraft}. We dive deeper into the benchmarking statistics, and identify possible data leakage in Appendix~\ref{subsec:app-problem-cd}.

\paragraph{Models underperform on the PoseBusters held-out test.}
In Table~\ref{tab:main}, we select competitive baselines on CrossDock, and run the sampling given their released codebases and checkpoints. 
Sampling efficiency measured in the ratio of successfully generated molecules (Connected) shows that autoregressive models severely degrade on these challenging targets.
Among non-autoregressive baselines, DecompDiff exhibits the most prominent degeneration in Vina affinities, while TargetDiff displays a notable drop in PB-Valid and ranks the least, and MolCRAFT shows significantly more strained structures reflected by the tail distribution of Strain Energy in Fig.~\ref{tab:strain_results}.
This suggests that PoseBusters can serve as a challenging OOD test for de novo design.

\paragraph{Our method is robust on the difficult PoseBusters test.} \ours~reliably generates physically valid conformations within the protein pocket with an impressive 56.1\% of cases with scRMSD $<$ 2\r{A}, comparable to the 59.4\% obtained by redocking the co-crystallized ligands, which indicates its ability to capture the interaction patterns and maintain binding pose consistency. 
We additionally calculate the Tanimoto similarity of interaction profiles between generated and ground-truth molecules via ProLIF \citep{bouysset2021prolif}. Table~\ref{tab:interaction_sim} shows that ours best matches the genuine interaction pattern of the co-crystallized structures.

Furthermore, considering that the accuracy of Vina docking tool evaluated on ground-truth complexes in fact places an upper bound, our result of 56.1\% suggests that Vina might be reaching its limit in evaluating the pose generation accuracy.
This brings us one step further towards ground truth-based evaluation, i.e., molecular docking.



\subsection{Molecular Docking}
\begin{figure*}[t]
    \centering
    \includegraphics[width=0.9\linewidth]{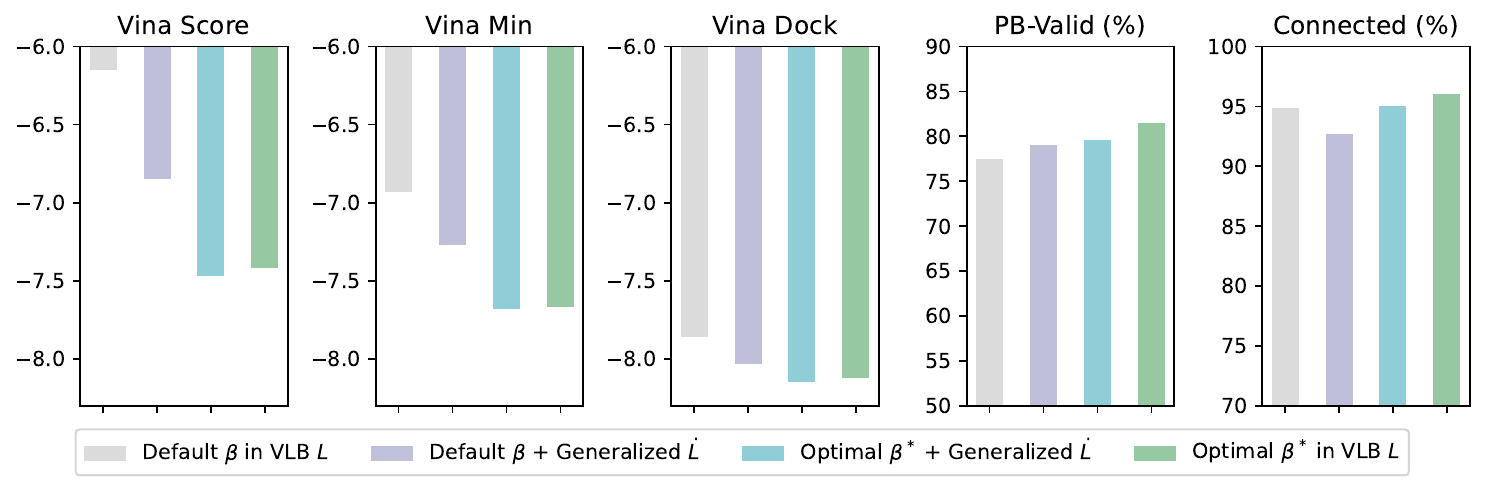}
    \caption{Ablation studies regarding loss objectives and noise schedules. Training with VLB (Eq.~\ref{eq:VLB}) as a line integral requires specific schedule $\boldsymbol\beta$ in training and sampling, while with generalized loss (Eq.~\ref{eq:generalized_loss}) as a surface integral it only requires $\boldsymbol\beta$ for test time.}
    \label{fig:ablation}
\end{figure*}

\begin{table*}[t]
\caption{Results of applying VOS to diffusion model (TargetDiff) on CrossDock. Top-1 results highlighted in \textbf{bold}. }
\centering
\label{tab:general_vos}
\resizebox{0.95\linewidth}{!}{%
\begin{tabular}{l|c|cc|cc|cc|c|c|c|c|c}
\toprule
\multirow{2}{*}{Methods} & PB-Valid$\dagger$ & \multicolumn{2}{c|}{Vina Score ($\downarrow$)} & \multicolumn{2}{c|}{Vina Min ($\downarrow$)} & \multicolumn{2}{c|}{Vina Dock ($\downarrow$)} & scRMSD & Energy                      & Connected$\dagger$ & QED & SA  \\
                         & Avg. ($\uparrow$)                 & Avg.                   & Med.                  & Avg.                  & Med.                 & Avg.                  & Med.                  & \textless 2 \r{A} ($\uparrow$) & Passed$\dagger$ ($\uparrow$) & Avg. ($\uparrow$)              & Avg.    & Avg.  \\ \midrule
TargetDiff        & 50.5\%   & -5.47          & -6.30          & -6.64        & -6.83        & -7.80         & -7.91         & 37.1\%             & 69.8\%        & 90.4\%    & 0.48 & 0.58 \\
TargetDiff + $\dot\gL$       & 53.7\%   & -6.27          & -6.31          & -6.82        & -6.78        & -7.87         & -7.90         & 36.8\%             & 70.3\%        & 89.2\%    & 0.50  & 0.62 \\
\rowcolor{gray!20}
TargetDiff + $\dot\gL$ + $\boldsymbol\beta^*$ (VOS) & \textbf{58.1\%} & \textbf{-6.46} & \textbf{-6.53} & \textbf{-7.04} & \textbf{-7.09} & \textbf{-8.04} & \textbf{-8.12} & \textbf{40.2\%}    & \textbf{73.2\%} & \textbf{93.4\%} & 0.49 & 0.59  \\ \bottomrule
\end{tabular}
}
\end{table*}

We note that the model trained with $\dot\gL$ in Eq.~\ref{eq:generalized_loss} is also available for multimarginal generative modeling \citep{pmlr-v235-campbell24a}, i.e., it can be used for molecular docking by an induced $P(\vr\mid \{\vh, \mA\}, \rvx_P)$.


For a deeper understanding of our model's ability in capturing genuine spatial interaction, we additionally evaluate its docking accuracy on PoseBusters V2 dataset with holo pocket residue structures specified, and report the RMSD results. While we did not specifically optimize our design w.r.t. molecular docking, Fig.~\ref{fig:main-dock} suggests that our method is a competitive SBDD model even compared with the local docking methods. 
This capability in recovering ground truth binding poses serves as an indicator of capturing true interactions, which is essential for designing bioactive molecules with potential biological efficacy. We believe that incorporating molecular docking as a subtask of SBDD also sheds light on the model's ability of interaction modeling.


\subsection{Ablation Studies}
We conduct ablation studies by sampling 10 molecules for each of 180 proteins in PoseBusters OOD test, showing the significance of our VOS method in Fig.~\ref{fig:ablation}.

\paragraph{Effect of generalized loss $\dot\gL$}
The generalized training objective boosts the generative model's ability in capturing spatial interactions, indicated by the improved Vina affinities especially the Vina Score that directly scores the generated poses. We attribute this gain to the fact that this loss objective forces the model to learn better-balanced probability path for multi-modality molecular data as shown in Fig.~\ref{fig:loss-decouple}, especially the previously missing ability to denoise continuous 3D positions with cleaner discrete 2D information on the region of higher-noise 3D and lower-noise 2D data.

\paragraph{Effect of optimal schedule $\boldsymbol\beta^*$}
When training under the generalized objective $\dot\gL$ and sampling under different noise schedules $\boldsymbol\beta$, it can be seen that optimal schedule $\boldsymbol\beta^*$ consistently achieves better performance than default noise schedule. Moreover, the optimal noise schedule can be further adopted in training under the VLB objective, yielding a slightly improved performance. This is due to the fact that VLB remains invariant as a line integral along the path determined by the optimal noise schedule function, yet the  denoising model effectively allocates more capacity in optimizing the VLB, leading to significantly faster convergence in about half the training time with generalized objective $\dot\gL$.


\subsection{Generality of VOS}
To demonstrate VOS's broader applicability, we have integrated it with the diffusion-based framework TargetDiff. We train for 140k steps following the default configuration with our generalized objective $\dot\gL$, and then derive the optimal schedule that proves to resemble the shape in Fig.~\ref{fig:surface}B. 

We sample 10 molecules per target on CrossDock and report the results in Table~\ref{tab:general_vos}, showing that VOS successfully enhances conformation quality for diffusion models as well, with generated poses achieving Vina Scores closely matching Vina Min values, indicating near-optimal realistic poses.

\section{Related Works}
\paragraph{Structure-based Drug Design (SBDD)}
SBDD generative models focus on the joint generation of discrete molecular topology and continuous conformation conditioned on protein-binding pockets. Recent approaches have centered on capturing the critical protein-ligand interactions for the generation of high-affinity molecules. Autoregressive models \citep{luo_3d_2022, peng_pocket2mol_2022, liu2022graphbp} generate molecules atom-by-atom while preserving geometric equivariance but are computationally expensive. Fragment-based methods \citep{powers2022fragmentbased, zhang_molecule_2023, lin2023d3fg} improve efficiency by generating motifs instead of atoms, though they often require post-processing to mitigate error accumulation.
Non-autoregressive models such as diffusion-based approaches \citep{schneuing_structure-based_2022, guan_3d_2023, guan_decompdiff_2023} and Bayesian Flow Networks (BFNs) \citep{qu2024molcraft} focus on full-atom generation, enabling scalability and improved controllability. 
Another line of works incorporates interaction context or property guidance signals such as binding affinity, to enhance the profile of generated ligands \citep{huang2024interactionbased, huang2024protein, qiu2024structure}. 
While significant progress has been made, challenges in conformation quality and interaction modeling remain central to advancing structure-based drug design \citep{harris2023benchmarking}.

\paragraph{Molecular Docking}
In the context of SBDD, molecular docking concerns predicting the 3D conformation given 2D molecular topology and the protein pocket structures. Traditional search-based local-docking approaches include AutoDock Vina \citep{eberhardt2021autodock}, Gold \citep{jones1997development}. Deep learning-based methods are divided into blind docking methods with holo protein structures specified, such as DiffDock \citep{corso2023diffdock}, and local docking methods such as DeepDock \citep{mendez2021geometric} and Uni-Mol \citep{zhou2023unimol}. These methods additionally rely on the RDKit-initialized molecular conformation as input.

\section{Conclusion}
This work addresses a critical challenge in Structure-Based Drug Design (SBDD) by introducing better-balanced dynamics in the generative process of multi-modalities. Equipped with an optimal noise schedule in terms of Variational Lower Bound (VLB), \ours~achieves a SOTA PoseBusters passing rate of 95.9\% on CrossDock with improved molecular geometry and interaction modeling, offering a promising step forward in the field of drug discovery.

\section*{Impact Statement}
This work makes a contribution to the field of computational drug discovery by improving the modeling of protein-ligand interactions, which are crucial for identifying therapeutic compounds. 
Our approach can be generalized to other tasks involving multimodal generative modeling such as material design. Ultimately, we hope this research could accelerate the discovery of bioactive compounds by enabling more accurate and reliable in silico design of protein-binding ligands, leading to more effective therapeutic interventions.

\section*{Acknowledgments}
This work is supported by the Natural Science Foundation of China (Grant No. 62376133) and sponsored by Beijing Nova Program (20240484682) and the Wuxi Research Institute of Applied Technologies, Tsinghua University (20242001120).
The authors would like to thank Jingjing Gong, Hanlin Wu and Zhilong Zhang for their helpful comments on this work.


\bibliography{example_paper}
\bibliographystyle{icml2025}

\newpage
\appendix
\onecolumn

\section{Details about the Proposed Method}

\subsection{Dynamic Programming for Optimal Schedules}

\begin{algorithm}[H]
\caption{Dynamic Programming for Optimal Path}\label{algo:dp}
\KwIn{Cost matrix \(\mC \in \R^{M\times M \times 2}\), step budget \(L\), step scale $K$, default noise schedule $\boldsymbol{\beta}(t)$}
\KwOut{Optimal \({path}\), minimal cumulative cost \(J^*\)}

\textbf{Initialization:}\\
Fit a smooth loss surface $\tilde{\mC}$ using B-spline interpolation.\;\\
\(J[:M, :M, :N+1] \gets \infty\), \(J[0, 0, 0] \gets 0\)\\
\(\text{prev}[:M, :M, :N+1, :2] \gets -1\)\\
\(\text{active\_states} \gets \{(0, 0, 0)\}\) \\
\(\text{size} \gets \frac{K}{1 + \|\nabla \tilde{\mC}\|}\), where \(\|\nabla \tilde{\mC}\| = \sqrt{(\frac{\partial \tilde{\mC}}{\partial x})^2 + (\frac{\partial \tilde{\mC}}{\partial y})^2}\)

\textbf{Dynamic Programming:}\\
\For{\(l \gets 0\) \textbf{to} \(L-1\)}{
    \(\text{states} \gets \emptyset\)\\
    \ForEach{\((x, y, l) \in \text{active\_states}\)}{
        \ForEach{\((\eps_x, \eps_y) \in \text{valid\_steps}(\text{size}[x, y])\)}{
            \(~~x_t, y_t \gets x+\eps_x, y+\eps_y\)\\
            \( \boldsymbol{\alpha} \gets [\frac{\beta_c(x_t)-\beta_c(x)}{\eps_x}, \frac{\beta_d(y_t)-\beta_d(y)}{\eps_y}] \)\\
            \(cost \gets J[x, y, l] + \boldsymbol{\alpha}\cdot \mC[x_t, y_t]\) \\
            \If{\(J[x_t, y_t, l+1] > cost\)}{
                \(~~J[x_t, y_t, l+1] \gets cost\)\\
                \(\text{prev}[x_t, y_t, l+1] \gets (x, y)\)\\
                \(\text{states} \gets \text{states} \cup \{(x_t, y_t, l+1)\}\)
            }
        }
    }
    \(\text{active\_states} \gets \text{states}\)
}

\(~~x, y, path \gets M-1, M-1, \emptyset\) \\
\(l \gets \argmin J[x, y, :]\)\\
\While{\(l \geq 0\)}{
    \(path.\text{append}((x, y))\)\\
    \((x, y) \gets \text{prev}[x, y, l]\), \(l \gets l - 1\)
}

\Return{\(path, \min J[M-1, M-1, :]\)}
\end{algorithm}

\paragraph{Convergence guarantee} 
As the grid resolution \(N \to \infty\), the discrete solution converges to the continuous VLB-optimal path \(\boldsymbol{\beta}^*(t)\). Since the VLB for a joint schedule \(\boldsymbol{\beta}(t)\) is a line integral over the model’s loss field in the 2D noise space (Appendix~\ref{subsec:app-vlb-line-integral}), and training with the generalized loss \(\dot{\gL}^\infty\) ensures the model’s predictions are accurate everywhere in this space, enabling the evaluation of VLBs for arbitrary paths \(\boldsymbol{\beta}(t)\) (Appendix~\ref{subsec:app-vlb-from-general}), the dynamic programming solution recovers the optimal path asymptotically, maximizing the VLB over the function space \(\gZ\).

\paragraph{Time complexity}
The computational complexity is $O(NM^2K)$, which does not add too much computational overhead and can be solved within a few minutes on a single CPU. Once solved, the optimal schedule can be adopted for test time, enabling the VLB-optimal generative process, where the multi-modalities effectively inform each other at the optimal accuracy level. However, in order to estimate the cost matrix over $M\times M$ discretized grid, the same number of evaluations of the generative model $\tilde{\rvx}_{\phi}$ is required, which scales quadratically to the discretization $M$ and involves considerable GPU computation.
Empirically, we found that the loss surface is smooth enough to allow for effective interpolation, therefore $M=20$ would yield an accurate enough estimate, taking less than 20 minutes.

\begin{table}[htbp]
    \centering
    \caption{Validation loss as the sum of scaled KL divergence at sample time for different scheduling.}
    \vskip 0.1in
    \resizebox{0.4\linewidth}{!}{
    \begin{tabular}{c|c|c|c}
        \toprule
        Schedule & Default & Optimal & Learned \\ \midrule
        Validation Loss & 5.51 & 4.23 & 5.39 \\ \bottomrule 
    \end{tabular}
    }
    \label{tab:VLB_comparison}
\end{table}

\subsection{Why Learning for Optimal Schedule Might Fail}

Given the obtained optimal schedule $\boldsymbol{\beta}^*(t)$, we can train the generative model with invariant optimal VLB (Eq.~\ref{eq:VLB}) more effectively, allocating the model capacity towards learning to denoise only along the selected path. We found this results in earlier convergence, taking around half the previous training time for generalized objective (Eq.~\ref{eq:generalized_loss}) with similar performance.

We would like to take one step further to see if we can directly optimize the joint noise schedule $\boldsymbol{\beta}(t)$ in generative training with the negative VLB objective in Eq.~\ref{eq:VLB}, i.e., optimizing the generative model and the noise schedule simultaneously through the same loss objective.

\begin{figure}
    \centering
    \includegraphics[width=0.6\linewidth]{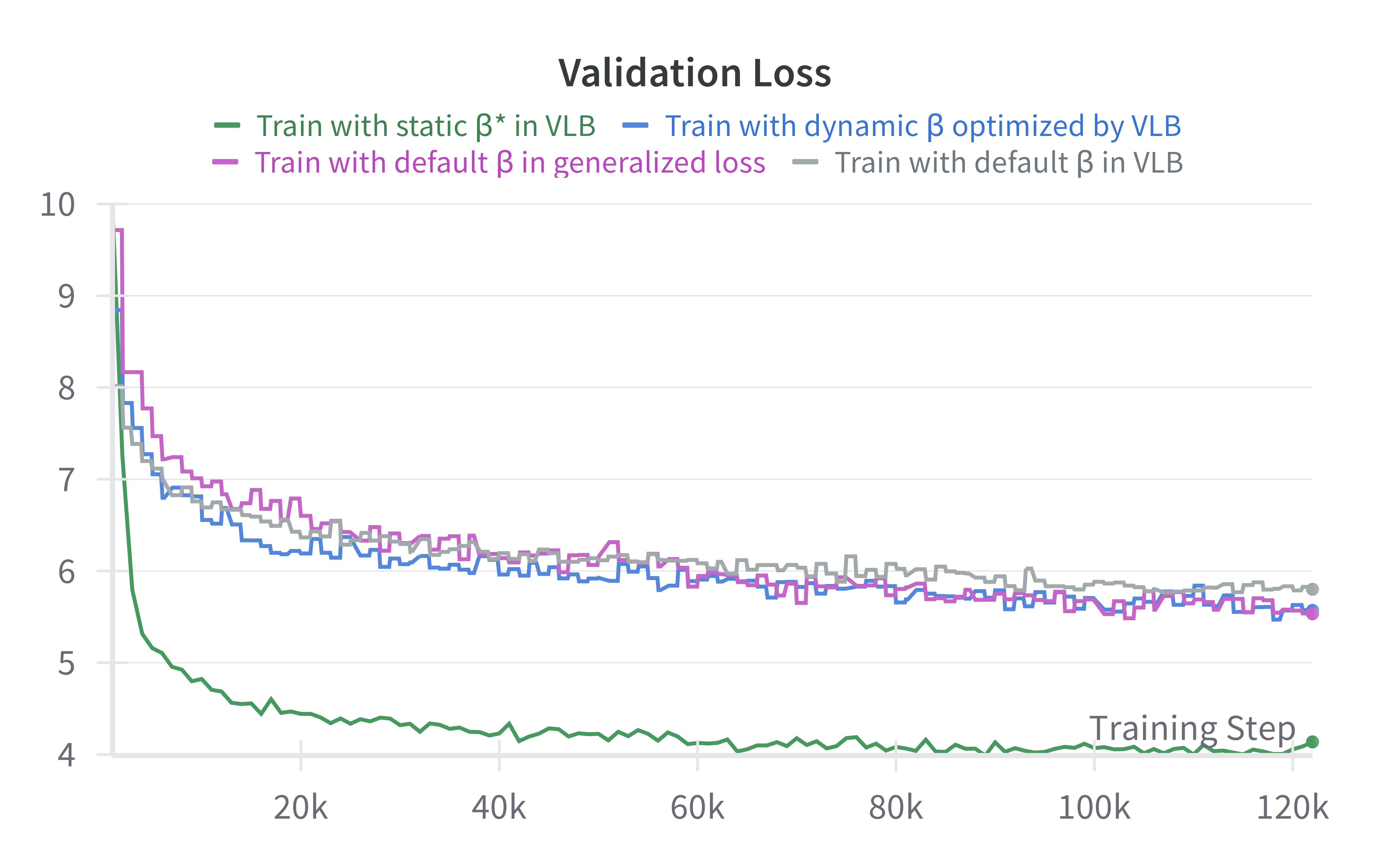}
    \caption{Validation losses for different configurations over the course of training. }
    \label{fig:val_loss_schedule}
\end{figure}

We parametrize the learnable time-rescaling function by
\begin{equation}
    f(t) = t + (1 - t)t \cdot \text{sigmoid}(h_\psi(t))
\end{equation}
where \( h_\psi(x) \) is a monotonic neural network which we choose to be a three-layered MLP with Softplus activation before the final output. This rescaling function is strictly monotonically increasing with endpoint constraints, i.e., $f(0)=0, f(1)=1$.

However, we found that the learned noise schedule does not change much from its initialized shape, and the corresponding VLB proves to be suboptimal in Table~\ref{tab:VLB_comparison}.
We hypothesize that there are several challenges in learning for the optimal schedule during training: 
(1) Large combinatorial space with limited exposure to possible noise schedules in the process of active learning.
Unlike single-modality models, where the range of noise levels is easily explored within the modality \citep{dieleman2022continuous}, multi-modality generative models can suffer from exposure to only a subset of the possible noise combinations. This incomplete exploration makes it challenging to optimize directly for the best schedule, as large regions of the design space remain unvisited.
It impacts the VLB estimation since the model $\tilde{\rvx}_\phi$ is still learning to adapt to the scaled noisy inputs during training, and is not invariant w.r.t. noisy input at different scales.
(2) Optimization difficulties. Directing taking the gradient w.r.t. $t_c, t_d$ introduces numerical instability, particularly when $\tilde{\beta_c}$ involves exponentiation, which can grow or shrink very rapidly for certain values of $t_c$, leading to vanishing or exploding gradients. Moreover, the VLB optimization is sensitive to the trajectory defined by the noise schedule, and the optimization landscape of may have many local minima, especially due to the coupling of noise schedules, which can make it difficult to find the global optimum.

Addressing these might require more sophisticated methods such as bilevel optimization, which we leave for future work.

\subsection{Sensitivity to Optimal Path Choice}

\paragraph{Effect of Interpolation}
We conduct additional experiments interpolating with a coefficient $c$ between identity time function $t_i$ ($c=0$) and our derived optimal time-rescaling functions $t_o$ ($c=1$), by setting time functions $t=c\cdot t_i + (1-c)t_o$. Our findings show a clear trend of improving performance as we move toward the optimal schedule.

\begin{table}[htbp]
\caption{Results of interpolating between identity time function ($c=0$) and optimal time-rescaling functions ($c=1$), where for each protein target 10 molecules are sampled.}
\vskip 0.1in
\centering
\label{tab:interpolation}
\resizebox{0.7\linewidth}{!}{%
\begin{tabular}{@{}l|cc|cc|c|c|c|c@{}}
\toprule
\multirow{2}{*}{$c$} & \multicolumn{2}{c|}{Vina Score ($\downarrow$)} & \multicolumn{2}{c|}{Vina Min ($\downarrow$)} & QED  & SA   & Connected         & PB-Valid          \\
                     & Avg.                   & Med.                  & Avg.                  & Med.                 & Avg. & Avg. & Avg. ($\uparrow$) & Avg. ($\uparrow$) \\ \midrule
CrossDock (ID)       & -6.36                  & -6.46                 & -6.71                 & -6.49                & 0.48 & 0.73 & -                 & 0.95              \\ \midrule
0 (Identity)        & -6.63                  & -6.94                 & -7.06                 & -7.06                & 0.55 & 0.76 & 0.95              & 0.95              \\
0.25                 & -6.66                  & -6.88                 & -7.02                 & -7.01                & 0.55 & 0.76 & 0.94              & 0.96              \\
0.5                  & -6.70                  & -6.82                 & -7.04                 & -6.94                & 0.55 & 0.76 & 0.95              & \textbf{0.97}     \\
0.75                 & -6.87                  & -6.97                 & -7.21                 & -7.11                & 0.55 & 0.76 & \textbf{0.96}     & 0.96              \\
1 (Optimal)         & \textbf{-6.92}         & \textbf{-7.02}        & \textbf{-7.23}        & \textbf{-7.18}       & 0.55 & 0.75 & \textbf{0.96}     & 0.95              \\ \midrule
PoseBusters (OOD)    & -7.06                  & -7.05                 & -7.50                 & -7.41                & 0.40 & 0.72 & -                 & 0.99              \\ \midrule
0 (Identity)        & -7.35                  & -7.38                 & -7.64                 & -7.48                & 0.48 & 0.73 & 0.94              & 0.79              \\
0.25                 & -7.42                  & -7.37                 & -7.66                 & -7.50                & 0.48 & 0.73 & 0.94              & 0.79              \\
0.5                  & -7.20                  & -7.36                 & -7.56                 & -7.51                & 0.48 & 0.73 & \textbf{0.95}     & 0.78              \\
0.75                 & -7.44                  & -7.43                 & -7.75                 & -7.54                & 0.48 & 0.73 & \textbf{0.95}     & 0.79              \\
1 (Optimal)         & \textbf{-7.52}         & \textbf{-7.52}        & \textbf{-7.79}        & \textbf{-7.65}       & 0.49 & 0.73 & \textbf{0.95}     & \textbf{0.80}     \\ \bottomrule
\end{tabular}
}
\end{table}

Table~\ref{tab:interpolation} shows a generally monotonic improvement in binding pose quality on both datasets as we move from identity to optimal scheduling functions, with the most significant gains occurring when $c \in [0.75, 1]$. Additionally, the case with $c=0$ also demonstrates the performance boost brought by our generalized training objective compared with the default objective (not shown in the table, please refer to Fig.~\ref{fig:ablation}). 

\paragraph{Effect of Accuracy Level $\boldsymbol\alpha$ as Scaling Factor}
The desired scaling factor $\boldsymbol{\alpha}$ ought to be the derivative of joint schedule function $\boldsymbol{\beta}$, which is not known before solving the search problem. To determine the appropriate accuracy level in practice, we have empirically found that we can actually approximate the accuracy by taking the derivative directly from the default schedule function $\boldsymbol{\tilde\beta}(t)=({\tilde\beta}_c'(t), {\tilde\beta}_d'(t))$, instead of taking the slope $\frac{\boldsymbol\beta(t) - \boldsymbol\beta(t-\eps)}{\eps}$, and this applies both to the experiments with BFN and diffusion models like TargetDiff. 
Note that by setting the accuracy level as such, the optimization objective no longer corresponds to the exact likelihood, but a rescaled sum of KL divergence terms that put more weight on the continuous variable. We hypothesize that this suggests a gap between likelihood estimation and sample quality, where the latter is more influenced by the 3D structure part.

\subsection{Model Architecture}
We employ the equivariant Graph Transformer architecture similar to \citet{guan_decompdiff_2023}, with a few modifications to attention score calculations designed to save memory. Denote node positions as $\mathbf{r} \in \mathbb{R}^{N\times 3}$, one-hot atom types as $\mathbf{h} \in \R^{N\times K_h}$, and one-hot bond types as $\mathbf{A} \in \R^{N\times(N-1)\times K_A}$, where $K_h$ and $K_A$ are the numbers of atom and bond types, respectively. We brief the model architecture below.

\paragraph{Heterogeneous Message Passing for Node Update}
The heterogeneous update for node representation consists of two parts: (1) a $K$-nearest neighbors graph $\mathcal{G}_K$ built for interactions within the protein-ligand complex, and (2) a fully connected ligand graph $\mathcal{G}_L$ within ligand atoms for ligand bond-based interactions.

\begin{equation*}
\Delta \mathbf{h}_{K,i} \leftarrow \sum_{j \in \mathcal{N}_K(i)} \phi_{h_K}(\mX_{kv}^K, \mX_{q}^K, \text{concat}(\mathbf{d}_{ij}, \mathbf{h}^{\text{edge}}_{ij})), \quad
\Delta\mathbf{h}_{L,i} \leftarrow \sum_{j \in \mathcal{N}_L(i)} \phi_{h_L}(\mX_{kv}^L, \mX_{q}^L, \mathbf{h}_{ij}^\text{bond}),
\end{equation*}
where the input features for complex graph $\mX_{kv}^K=\text{concat}(\mathbf{h}_i, \mathbf{h}_j, \mathbf{h}^{\text{edge}})$, $\mX_q^K= \text{concat}(\mathbf{h}_i, \mathbf{h}^{\text{edge}})$, \(\mathbf{d}_{ij}=\|\mathbf{r}_i - \mathbf{r}_j\|\) is the pairwise distance, and \(\mathbf{h}^{\text{edge}}\) is the one-hot encoding for edge types between protein-ligand atoms (protein-protein, protein-ligand, ligand-protein or ligand-ligand).
For ligand graph, the input features is defined as $\mX_{kv}^L = \text{concat}(\mathbf{h}_i, \mathbf{h}_j, \mathbf{h}_{\text{bond}})$, $\mX_q^L = \text{concat}(\mathbf{h}_i, \mathbf{h}_{\text{bond}})$, where $\mathbf{h}_{\text{bond}}$ denotes the flattened bond embeddings.

The node-level representation is updated by aggregating the heterogeneous messages:
\begin{equation*}
    \mathbf{h}_i \leftarrow \mathbf{h}_i + \text{FFN}\big(\mathbf{h}_i + (\Delta\mathbf{h}_{K,i} + \Delta\mathbf{h}_{L,i})\mW_h\big).
\end{equation*}

\paragraph{Intra-Ligand Message Passing for Bond Update}
The bond representation is updated by a directional message passing:
\begin{equation*}
    \mathbf{h}_{ji}^\text{bond} \leftarrow \sum_{k \in \mathcal{N}_L(j) \setminus \{i\}} \phi_E(\mathbf{X}_{kv}^E, \mathbf{X}_q^E, \sigma(\mathbf{d}_{ji})).
\end{equation*}
where the input features are defined as $\mathbf{X}_{kv}^E = \text{concat}(\mathbf{h}_j, \mathbf{h}_k, \mathbf{m}_{kj})$, $\mathbf{X}_q^E = \text{concat}(\mathbf{h}^{\text{bond}}_{kj}, \sigma(\mathbf{d}_{kj}), \sigma(\mathbf{d}_{ji}), \mathbf{a}_{ijk}, \mathbf{h}_k, \mathbf{h}_j)$, and $\mathbf{X}_{q}^E = \text{concat}(\mathbf{h}^{\text{bond}}_{ji}, \mathbf{h}_i)$. $\sigma$ is the Gaussian smearing function applied to the pairwise distance, and $\mathbf{a}$ is the angular encoding for the bond triplets.

\paragraph{Heterogeneous Message Passing for Position Updates}
Similarly, the positions are updated from heterogeneous messages:
\begin{equation*}
    \Delta\mathbf{r}_{K,i} \leftarrow \sum_{j \in \mathcal{N}_K(i)} (\mathbf{r}_j - \mathbf{r}_i)\odot\phi_{r_K}(\mX_{kv}^K, \mX_{q}^K, \text{concat}(\mathbf{d}_{ij}, \mathbf{h}^{\text{edge}}_{ij})), \quad
\Delta\mathbf{r}_{L,i} \leftarrow \sum_{j \in \mathcal{N}_L(i)} (\mathbf{r}_j - \mathbf{r}_i)\phi_{r_L}(\mX_{kv}^L, \mX_{q}^L, \mathbf{h}_{ij}^\text{bond})
\end{equation*}
given the updated node features $ \mathbf{h}_{ij}$ and bond features $\mathbf{h}_{ij}^\text{bond}$. The final position is updated by
\begin{equation*}
\mathbf{r}_i \leftarrow \mathbf{r}_i + (\Delta\mathbf{r}_{K,i} + \Delta\mathbf{r}_{L,i}) \cdot \mathbf{1}_{\text{mol}}
\end{equation*}
where $\bold{1}_{\text{mol}}$ is the boolean mask for ligand atoms, as the protein atoms need to be fixed to ensure equivariance.

\paragraph{Attention Mechanism}
The function $\phi$ is implemented as a multi-headed attention mechanism. The query, key, value are obtained by passing input features through linear layer without bias to save memory:
\[
\mathbf{K}=\mX_{kv}\mW_k,\quad \mathbf{V} = \mX_{kv}\mW_v,\quad \mathbf{Q}=\mX_q\mW_q.
\]
The attention scores are computed using:
\[
\boldsymbol{\alpha}_{ij} = \frac{\mathbf{q}_i (\mathbf{k}_j \odot \mathbf{w}_{ij})}{\sqrt{d_h}}, \quad \mathbf{w}_{ij}  = \tanh(\mathbf{e}_{ij}\mathbf{W}_{\text{edge}} ),
\]
where \(d_h\) is the head dimension, \(\mathbf{w}_{ij}\) is the edge-specific weights modulating the contribution of distant neighbors, and $\odot$ denotes Hadamard product.

\subsection{Implementation Details}
\paragraph{Training and Inference}
We use Adam optimizer with learning rate 5e-4, batch size of 16, and fit the model with 3.1 million parameters on one NVIDIA 80GB A100 GPU. We set $\beta_1 = 1.5$ for discrete atom types and bond types, $\sigma_1 = 0.05$ for atom coordinates. The training converges in 200K steps (around 24 hours). 
For inference, we use the exponential moving average of the weights from training that is updated at every optimization step with a decay factor of 0.999. We run inference with 100 sampling steps with the same variance reduction sampling strategy as \citet{qu2024molcraft}.

\paragraph{Hyperparameters for Network}
We set the network to be kNN graphs with $k=32$, $N=9$ layers with $d=128$ hidden dimension, 16-headed attention, and dropout rate 0.1.

\paragraph{Featurization}
We describe our atom-level and edge-level featurization for the protein-ligand complexes. At the atom-level, each protein atom is represented with a one-hot element indicator (H, C, N, O, S, Se), a 20-dimensional one-hot amino acid type indicator, and a 1-dimensional backbone flag. Ligand atoms are featurized with a one-hot element indicator (C, N, O, F, P, S, Cl) coupled with an aromaticity flag. 
For edge-level featurzation in the heterogeneous protein-ligand graph, edge types are encoded as a 4-dimensional one-hot vector indicating whether the edge is between ligand atoms, protein atoms, ligand-protein atoms, or protein-ligand atoms. For the ligand graph, bonds are represented with a 4-dimensional one-hot bond type vector (non-bond, single, double, triple), where aromatic bonds are kekulized using RDKit. 

\section{Proof}
\subsection{Derivation of Eq.~\ref{eq:kl_bfn} as single-modality VLB (Eq.~\ref{eq:vlb_bfn})}\label{subsec:app-vlb}
We begin with the negative Variational Lower Bound (VLB) in Eq.~\ref{eq:vlb_bfn}:
\begin{align}
    \gL_{\text{VLB}}(\rvx)
    \;=\;
    - \E_{q(\rvy_{1:n}\mid\rvx)}\!\bigl[\log p(\rvx\mid\rvy_{1:n})
    \;+\;
    \sum_{i=1}^n
    \KL\!\Bigl(q(\rvy_{i}\mid\rvx) \;\big\Vert\; p(\rvy_{i}\mid \rvy_{1:i-1})\Bigr)\bigr]. \nonumber
\end{align}

In further analysis, we focus on the treatment of the second term on the right-hand side. For convenience in the subsequent derivation, we define:
\begin{align}\label{eq:Ln-KL-sum}
    \gL^n(\rvx) \;&=\;
    \E_{q(\rvy_{1:n}\mid\rvx)}
    \Bigl[
        \sum_{i=1}^n
        \KL\!\bigl(q(\rvy_{i}\mid\rvx) \;\Vert\; p(\rvy_{i}\mid \rvy_{1:i-1}, \rvx)\bigr)
    \Bigr] \nonumber \\
    \;&=\;
    n \;\E_{i\sim U(1,n), q(\rvy_{1:n}\mid\rvx)}
    \Bigl[
    \KL\!\bigl(q(\rvy_{i}\mid\rvx) \;\Vert\; p(\rvy_{i}\mid \rvy_{1:i-1}, \rvx)\bigr)
    \Bigr]
\end{align}

Next, to specify the conditional distribution $p(\rvy_{i}\mid \rvy_{1:i-1}, \rvx)$, we utilize $p(\rvy_i \mid \rvy_{1:i-1}, \rvx)  \;=\; \frac{p(\rvy_{1:i} \mid \rvx)}{p(\rvy_{1:i-1} \mid \rvx)}$, and incorporate the following decomposition:
\begin{equation}
    p(\rvy_{1:i} \mid \rvx)
    \;=\;
    \prod_{j=1}^i 
    \E_{p_F(\vtheta_{\,j-1}\mid \rvx, t_{j-1})}\!
    \bigl[p_R(\rvy_j \mid \vtheta_{\,j-1}, t_{j-1})\bigr]
\end{equation}
It follows that:
\begin{equation}
    p(\rvy_i \mid \rvy_{1:i-1}, \rvx)
    \;=\;
    \E_{p_F(\vtheta_{\,i-1}\mid \rvx, t_{\,i-1})}
    \Bigl[p_R(\rvy_i \mid \vtheta_{\,i-1}, t_{\,i-1})\Bigr]
\end{equation}

Substituting this result into Eq.~\ref{eq:Ln-KL-sum}, we obtain:
\begin{align}\label{eq:Ln-KL-theta}
    \gL^n(\rvx)
    \;=\;
    n \;\E_{i \sim U(1,n), \vtheta \sim p_F(\vtheta \mid \rvx, t_{i-1})}
    \Bigl[
        \KL\!\bigl(q(\rvy_i \mid \rvx) \;\Vert\; p_R(\rvy_i \mid \vtheta, t_{i-1})\bigr)
    \Bigr]
\end{align}

Given our parametrization with neural network, we can further replace $p_R(\rvy_i \mid \vtheta, t_{i-1})$ with $q\bigl(\rvy_i \mid \tilde{\rvx}_\phi(\vtheta, t)\bigr)$:
\begin{align}\label{eq:Ln-KL-q}
    \gL^n(\rvx)
    \;=\;
    n \;\E_{i \sim U(1,n), \vtheta \sim p_F(\vtheta \mid \rvx, t_{i-1})}
    \Bigl[
        \KL\!\bigl(q(\rvy_i \mid \rvx) \;\Vert\; q(\rvy_i \mid \tilde{\rvx}_\phi(\vtheta, t))\bigr)
    \Bigr]
\end{align}

Following \citet{graves2023bayesian}, different noising processes are specified for $q$ for different modalities. For continuous data, $q(\rvy \mid \rvx) = \gN(\rvx, \alpha^{-1}\mI)$. For discrete data, $q(\rvy \mid \rvx) = \gN(\alpha(K{\rve_\rvx} - \mathbf{1}), \alpha K\mI)$, where we define $\rvx := \sqrt{K} \rve_\rvx$. 
Denoting $\gL^{\infty}(\rvx) 
    \;=\; 
    \lim_{n \to \infty} \gL^n(\rvx), 
    \alpha(t,\epsilon) = \beta(t) - \beta\bigl(t-\epsilon\bigr)$, we have
\begin{align}
    \gL^\infty(\rvx) & = \lim_{\epsilon \to 0} \frac{1}{\epsilon} \; \E_{i\sim U(1,n), p_F(\vtheta \mid x, t_{i-1})} \KL(q(\rvy_{i}\mid\rvx) ~\Vert~ q(\rvy_i \mid \tilde{\rvx}_\phi(\vtheta, t))) \nonumber \\
    & = \lim_{\epsilon \to 0} \E_{i\sim U(1,n), p_F(\vtheta \mid x, t_{i-1})} \frac{\alpha(t, \epsilon)}{2\epsilon} \Vert \rvx - \tilde{\rvx}_\phi(\vtheta, t) \Vert^2 \nonumber \\ 
    & = \frac{1}{2} \; \E_{i\sim U(1,n), p_F(\vtheta \mid x, t_{i-1})} \beta'(t)\Vert \rvx - \tilde{\rvx}_\phi(\vtheta, t)\Vert^2
\end{align}
which establishes the fact that Eq.~\ref{eq:kl_bfn} corresponds to the single-modality VLB.

\subsection{Equivalence between Eq.~\ref{eq:kl_bfn} and Eq.~\ref{eq:kl_beta}}
Rewrite the expectation in the above equation into an integral form and make the substitution $u=\beta(t)$. This substitution can be transformed as follows: $u=\beta(t) \Rightarrow du = \beta'(t) dt \Rightarrow dt=\frac{du}{\beta'(t)}$, and we directly rewrite $\tilde{\rvx}_\phi(\vtheta, t)$ as $\tilde{\rvx}_\phi(\vtheta, u)$.

\begin{align}
    \gL^\infty(\rvx) & = \frac{1}{2}\int_0^1 \int_{p_F(\vtheta \mid x, t)} \beta'(t) \Vert \rvx - \tilde{\rvx}_\phi(\vtheta, t)\Vert^2 d\vtheta \; dt \nonumber \\
    & = \frac{1}{2}\int_{\beta(0)}^{\beta(1)} \int_{p_F(\vtheta \mid x, u)} \beta'(\beta^{-1}(u)) \Vert \rvx - \tilde{\rvx}_\phi(\vtheta, u)\Vert^2 \frac{du}{\beta'(\beta^{-1}(u))} d\vtheta \nonumber \\
    & = \frac{1}{2}\int_{\beta(0)}^{\beta(1)} \int_{p_F(\vtheta \mid x, u)}  \Vert \rvx - \tilde{\rvx}_\phi(\vtheta, u)\Vert^2 du \; d\vtheta \nonumber \\
    & = \frac{1}{2}\int_{\beta(0)}^{\beta(1)} \E_{p_F(\vtheta \mid x, u)} \Vert \rvx - \tilde{\rvx}_\phi(\vtheta, u)\Vert^2 du
\end{align}

which gives the form of Eq.~\ref{eq:kl_beta}.

\subsection{Equivalence between Eq.~\ref{eq:generalized_loss} and \ref{eq:generalized_loss_for_t}}
Recall that the generalized loss over $\boldsymbol{\beta}\in\gZ$ is:
\begin{equation*}
    \gL^\infty(\rvx) =  \frac{1}{2}\int_{\beta_c, \beta_d}\underset{p_F(\vtheta | \rvx; \beta)}{\E}\Vert \rvx - \tilde{\rvx}_\phi(\vtheta, \boldsymbol{\beta})\Vert^2 d\boldsymbol{\beta}.
\end{equation*}
and the function space $\gZ$ is reparameterized into a product of uniform distributions over $t_c, t_d$, yielding
\begin{align*}
\dot{\gL}^\infty(\rvx) = \frac{1}{2}\int_{0}^{1}\int_{0}^{1}\underset{p_F(\vtheta | \rvx; \boldsymbol{t})}{\E} \Vert \rvx - \tilde{\rvx}_\phi(\vtheta, \boldsymbol{t})\Vert^2 dt_c dt_d,
\end{align*}
this is equivalent to implicitly sampling $f(t), g(t)$, covering all possible time-rescaling functions as stated in Remark~\ref{rem:time_decouple}. 

The joint distribution over $(\beta_c, \beta_d)$ is thus 
\begin{align}
p(\beta_c, \beta_d) = p(f, g) = U(0,1) \times U(0,1),
\end{align}
where \(\beta_c(t) = \tilde{\beta}_c(f(t))\) and \(\beta_d(t) = \tilde{\beta}_d(g(t))\).

Therefore, we see that the time-rescaling functions $f(t), g(t)$ act as latent variables defining the coupling of $\beta_c, \beta_t$. Integrating over $(t_c, t_d)$ marginalizes over all possible $\beta$, ensuring the model is trained to denoise all possible combinations of noise levels.
By demonstrating this equivalence, we reframe the invariant objective described in Eq.~\ref{eq:generalized_loss} as Eq.~\ref{eq:generalized_loss_for_t}.

\subsection{Derivation of Eq.~\ref{eq:VLB} as multi-modality VLB}\label{subsec:app-vlb-line-integral}
To rigorously prove that the VLB corresponds to a line integral along a 1D submanifold (path) in the 2D noise schedule space, recall that the joint noise schedule be a parameterized path in the joint function space $\gZ$:
\[
\boldsymbol{\beta}(t) = \big(\beta_c(t), \beta_d(t)\big), \quad t \in [0, 1],
\]
where \(\beta_c(t), \beta_d(t)\) are monotonically increasing functions with fixed endpoints:
\[
\beta_c(0) = \beta_{c,0}, \quad \beta_c(1) = \beta_{c,1}, \quad \beta_d(0) = \beta_{d,0}, \quad \beta_d(1) = \beta_{d,1}.
\]

Following Appendix~\ref{subsec:app-vlb} and the factorized nature of $q$, the VLB for a joint schedule \(\boldsymbol{\beta}(t)\) and multi-modality $\rvx$ is:
\[
\gL_{\text{VLB}}(\rvx) = \frac{1}{2}\mathbb{E}_{p_F(\vtheta \mid \rvx; t)} \int_0^1 \left[\beta_c'(t) \Vert \rvx_c - \tilde{\rvx}_{\phi,c}(\vtheta, \boldsymbol{\beta}(t)) \Vert^2 + \beta_d'(t) \Vert \rvx_d - \tilde{\rvx}_{\phi,d}(\vtheta, \boldsymbol{\beta}(t)) \Vert^2 \right] dt,
\]
where \(\tilde{\rvx}_\phi = (\tilde{\rvx}_{\phi,c}, \tilde{\rvx}_{\phi,d})\) is the denoising model output for continuous (\(c\)) and discrete (\(d\)) modalities.

The VLB can be interpreted as a line integral over the trajectory \(\boldsymbol{\beta}(t) \in \gZ\). Define the vector field \(\mathbf{F}(\beta_c, \beta_d)\) as:
\[
\mathbf{F}(\beta_c, \beta_d) = \begin{pmatrix} \Vert \rvx_c - \tilde{\rvx}_{\phi,c}(\vtheta, \boldsymbol{\beta}) \Vert^2 \\ \Vert \rvx_d - \tilde{\rvx}_{\phi,d}(\vtheta, \boldsymbol{\beta}) \Vert^2 \end{pmatrix}.
\]
The VLB then becomes:
\[
\gL_{\text{VLB}}(\rvx) = \frac{1}{2}\mathbb{E}_{\vtheta} \int_{\boldsymbol{\beta}(t)} \mathbf{F}(\beta_c, \beta_d) \cdot d\boldsymbol{\beta},
\]
where \(d\boldsymbol{\beta} = (\beta_c'(t) dt, \beta_d'(t) dt)\) is the differential vector along the path \(\boldsymbol{\beta}(t)\), and $\cdot$ denotes inner product.
By redefining $\tilde\rvx_\phi(\vtheta, \boldsymbol{\beta}) := \rmF(\beta_c, \beta_d)$, we obtain Eq.~\ref{eq:VLB}.

\subsection{Derivation of Proposition~\ref{prop:vlb}}\label{subsec:app-vlb-from-general}

\begin{proof}
Since the model $\tilde{\rvx}_\phi(\vtheta, \boldsymbol{t})$ is trained by the generalized loss $\dot{\gL}^\infty$, which is equivalent to integrating the scalar field \(\Vert \rvx - \tilde{\rvx}_\phi \Vert^2\) over the entire function space $\gZ$, it is exposed to all pairs $(\beta_c, \beta_d) \in \gZ$ within the region $[\tilde\beta_c(0), \tilde\beta_c(1)] \times [\tilde\beta_d(0), \tilde\beta_d(1)]$, with $\tilde\beta_c, \tilde\beta_d$ defined in Eq.~\ref{eq:beta}.
Therefore, the model achieves minimal prediction error everywhere in the joint noise space, allowing accurate computation for any specific path $\boldsymbol{\beta}(t)$.

Following Appendix~\ref{subsec:app-vlb-line-integral}, the line integral $\gL^\infty(\rvx)$ corresponds to the VLB for a given path $\boldsymbol{\beta}(t)$. Since the integral is equivalent to restricting the generalized loss $\dot{\gL}^\infty(\rvx)$ to the 1D manifold defined by $\boldsymbol{\beta}(t)$, and the model's predictions are accurate along any submanifold within the space, evaluating $\gL^\infty(\rvx)$ for a specific $\boldsymbol{\beta}(t)$ uses the model’s pre-optimized predictions at each point on the path, thereby yielding a valid VLB estimate. This establishes that the generalized training enables estimating VLBs for arbitrary joint schedules $\boldsymbol{\beta} \in \gZ$. 
\end{proof}

\section{Details on Benchmarking}

\subsection{Problems with CrossDock}\label{subsec:app-problem-cd}
\begin{figure}
    \centering
    \includegraphics[width=0.8\linewidth]{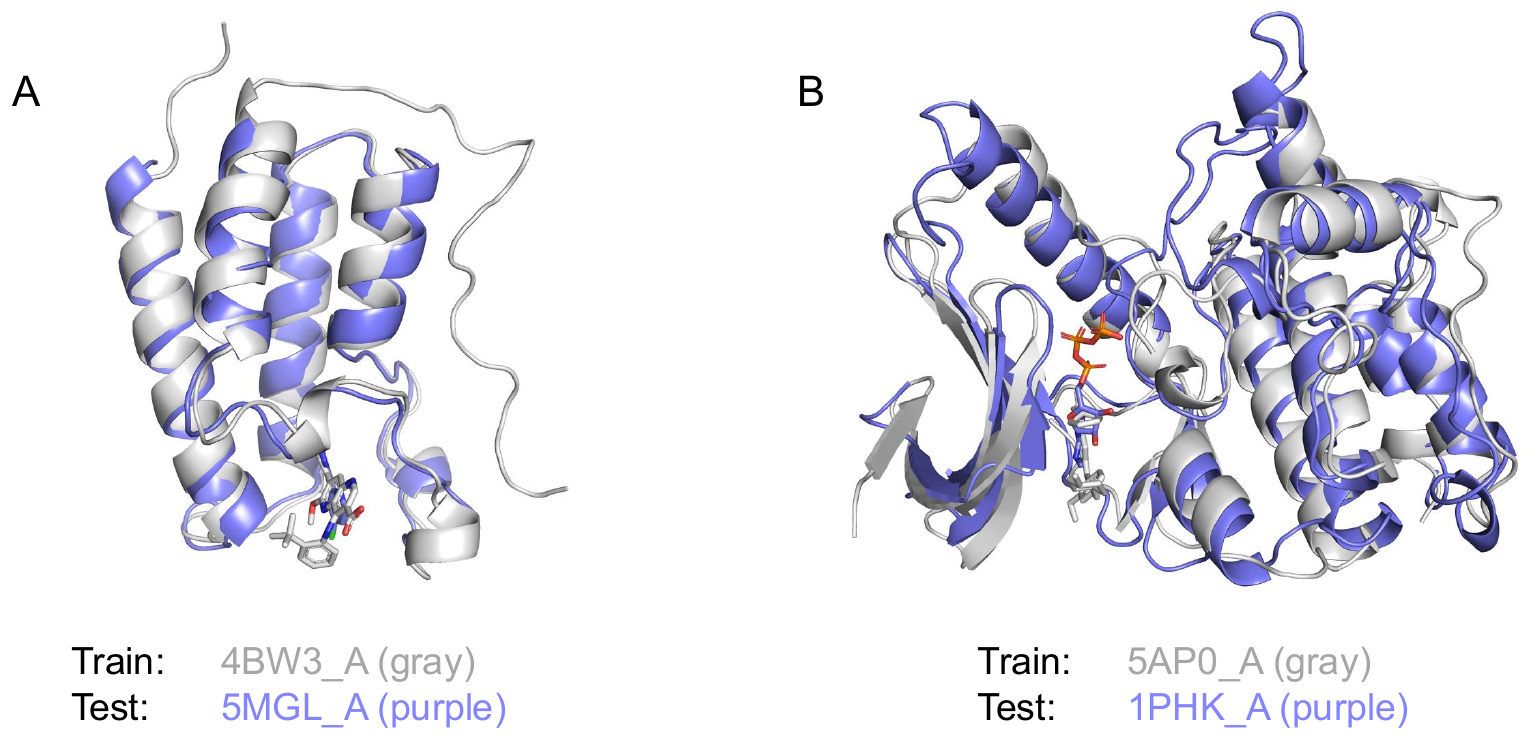}
    \caption{Structural alignment of randomly selected cases where training sequence and test sequence exhibit $>$ 30\% identity, showing nearly identical protein structures as well as ligand binding pockets.}
    \label{fig:mmseq-crossdock}
\end{figure}

\begin{table}[htbp]
\caption{Test protein chains that exhibit a sequence identity higher than 30\% to any chain of training sequences. For those with more than 10 overlapping training chains, we randomly select 10 items to display, and report the total number of unique overlapping chains. Each item is characterized by its PDB ID and chain ID.}
\label{tab:problem-cd}
\vskip 0.1in
\centering
\resizebox{0.8\columnwidth}{!}{%
\begin{tabular}{@{}lll@{}}
\toprule
Test PDB & \#Overlap & Training PDB(s) with Sequence Identity \textgreater 30\%                                 \\ \midrule
1PHK\_A  & 104       & 2WQO\_A, 3TKU\_B, 3WYY\_A, 3WYX\_A, 4C4E\_A, 4BHZ\_A, 2XKD\_A, 4ZEG\_A, 4CV8\_A, 5EI8\_A \\
1UMD\_B  & 1         & 1NI4\_B                                                                                  \\
5NGZ\_A  & 5         & 5F6Y\_A, 5F6W\_A, 5F6D\_A, 5F6E\_A, 5F6U\_A                                              \\
5L1V\_A  & 5         & 4DNJ\_A, 1Z8Q\_A, 1Z8O\_A, 1EGY\_A, 1JIP\_A                                              \\
1FMC\_B  & 8         & 5L7T\_A, 5ICS\_C, 5ICM\_A, 5L7W\_A, 5HS6\_A, 5EN4\_A, 5L7Y\_A, 5JS6\_A                   \\
4PXZ\_A  & 4         & 4Z34\_A, 4Z35\_A, 5WIV\_A, 4Z36\_A                                                       \\
3HY9\_B  & 1         & 1R55\_A                                                                                  \\
4AZF\_A  & 6         & 5J1V\_B, 5X8I\_B, 2B9J\_A, 1Z57\_A, 2VAG\_A, 5J1W\_C                                     \\
3PDH\_A  & 4         & 1YC2\_D, 1YC2\_B, 1S7G\_B, 1S7G\_C                                                       \\
5MGL\_A  & 127       & 5UVW\_A, 5D0C\_A, 4O7B\_A, 5KU3\_A, 5DX4\_A, 4A9F\_A, 5CS8\_A, 5D3L\_A, 4O7C\_A, 5E0R\_A \\
4XLI\_B  & 43        & 4OTF\_A, 4Y93\_A, 4Y95\_A, 5JRS\_A, 3GEN\_A, 5P9H\_A, 4NWM\_A, 4ZLY\_A, 3PIY\_A, 4RFY\_A \\
4AUA\_A  & 75        & 2R3Q\_A, 3EZR\_A, 3TN8\_A, 5FGK\_A, 2R3M\_A, 5IDN\_A, 2YIY\_A, 5HBJ\_A, 4CFU\_A, 3DOG\_A \\
5I0B\_A  & 100       & 5OPB\_A, 5OPR\_A, 2YM4\_A, 2C3K\_A, 3F69\_B, 4FTT\_A, 4FTO\_A, 3TKH\_A, 2BRG\_A, 2YDK\_A \\
4F1M\_A  & 6         & 5LPV\_A, 5LPW\_A, 5LPY\_A, 4OH4\_B, 4Q5J\_A, 5LPB\_A                                     \\
2CY0\_A  & 10        & 3PHJ\_A, 4FQ8\_B, 3PHH\_A, 3PGJ\_C, 3DON\_A, 4FOS\_A, 3PHG\_A, 3DOO\_A, 4FR5\_B, 4FPX\_A \\ \bottomrule
\end{tabular}%
}
\end{table}

The CrossDock dataset, first processed by \citet{luo_3d_2022}, is a commonly used benchmark in the SBDD field. \citet{luo_3d_2022} claimed to have employed the MMseqs2 method to filter the test set with a 30\% sequence similarity threshold. However, upon rigorous examination, we found that the test set still contains proteins with sequence similarity higher than 30\%. 

Table~\ref{tab:problem-cd} summarizes the calculated statistics for these similar proteins, where we only show the randomly sampled 10 PDB IDs when there are more than 10 sequences that exceed $>$ 30\% sequence identity. 
Fig.~\ref{fig:mmseq-crossdock} illustrates structural alignments between some random proteins in the test and training sets that exhibit high sequence similarity. The near-identical overlap suggests that the dataset split may not be challenging enough. 
To address these potential problems with CrossDock evaluation, we propose using PoseBusters \citep{buttenschoen2024posebusters} as a held-out test set.

\subsection{Curation of the held-out PoseBusters test set}
The PoseBusters Benchmark \citep{buttenschoen2024posebusters} set exclusively contains complexes released since 2021. In contrast, the CrossDock dataset comprises data collected from the PDB before 2020. This makes PoseBusters a challenging time-split dataset, ideal for evaluating the generalizability of models to real-world, unseen scenarios. We further apply MMseqs clustering \cite{steinegger2017mmseqs2}, and filter any test protein with any chain that has $>$ 30\% sequence identity threshold to CrossDock training sequences. This process leaves us with 180 data points, making most of the baselines directly available for the curated held-out test.
Additionally, the protein-ligand complex structures in CrossDock are mainly generated using docking software, which inevitably introduces noisy poses \citep{francoeur2020three}. In contrast, PoseBuster contains real-world crystal structures resolved from wet-lab experiments, making it suitable for evaluating the SBDD model’s ability to capture genuine molecular interactions accurately.

\subsection{Experimental Setup}\label{subsec:app-exp-setup}
\paragraph{Baselines}
We provide a brief overview of all SBDD baselines as follows:
\begin{itemize}
    \item \textbf{Autoregressive methods:} AR \citep{luo_3d_2022} utilizes MCMC sampling to reconstruct molecules atom-by-atom based on voxel-wise density predictions. Pocket2Mol \citep{peng_pocket2mol_2022} generates molecules atom-by-atom with bonds using an E(3)-equivariant network, predicting frontier atoms to improve sampling efficiency. FLAG \citep{zhang_molecule_2023} is a fragment-based model that assembles molecular fragments by predicting their positions and torsion angles.
    \item \textbf{Diffusion:} DiffSBDD \citep{schneuing_structure-based_2022} constructs an E(3)-equivariant continuous diffusion model for full-atom generation, applying noise to both atom types and coordinates, while TargetDiff \citep{guan_3d_2023} adopts a hybrid diffusion process to separately handle continuous coordinates and discrete atom types.
    DecompDiff \citep{guan_decompdiff_2023} incorporates chemical priors by decomposing molecules into scaffolds and contact arms.
    \item \textbf{BFN:} MolCRAFT \citep{qu2024molcraft} uses Bayesian Flow Network (BFN) with advanced variance reduction sampling technique, demonstrating notable improvements over diffusion-based models in molecular design.
\end{itemize}

\paragraph{Metrics}
We evaluate the generated molecules using the following commonly adopted metrics:

\begin{itemize}
    \item \textbf{Affinity Metrics} are calculated using AutoDock Vina \citep{eberhardt2021autodock}, these include \emph{Vina Score} as the raw binding energy of a molecular pose in the pocket, \emph{Vina Min} as the binding energy after local energy minimization of the molecular pose, and \emph{Vina Dock} as the lowest binding energy obtained after an extended search for the optimal pose.
    \item \textbf{Molecular Properties} including \textbf{QED} for drug-likeness and \textbf{SA} (synthetic accessibility score) are calculated using RDKit. They are desired to fall within reasonable ranges.
    \item \textbf{Connected Ratio} is the percentage of fully connected molecules.
    \item \textbf{Diversity} assesses the variety of generated molecules for each binding site by averaging Tanimoto similarity over Morgan fingerprints across all test proteins, following \citet{luo_3d_2022}. It is worth noting that this is not necessarily the higher the better, as the desirable bioactive compounds against a protein target often cluster in the molecular space.
    \item \textbf{Key Interactions} such as the formation of critical non-covalent interactions with protein binding sites, calculated using ProLIF \citep{bouysset2021prolif}.
    \item \textbf{Strain Energy} reflects the internal energy of generated poses, indicating pose quality by \citet{harris2023benchmarking}.
    \item \textbf{RMSD} reports the percentage of molecules where the RMSD between generated poses and ground-truth or Vina redocked poses is within 2 \r{A}, indicating consistent binding modes. For better differentiation, we refer to the latter as self-consistency RMSD (scRMSD).
\end{itemize}

\section{More Evaluation Results}\label{sec:app-exp}

\begin{figure*}[t]
    \centering
    \includegraphics[width=\linewidth]{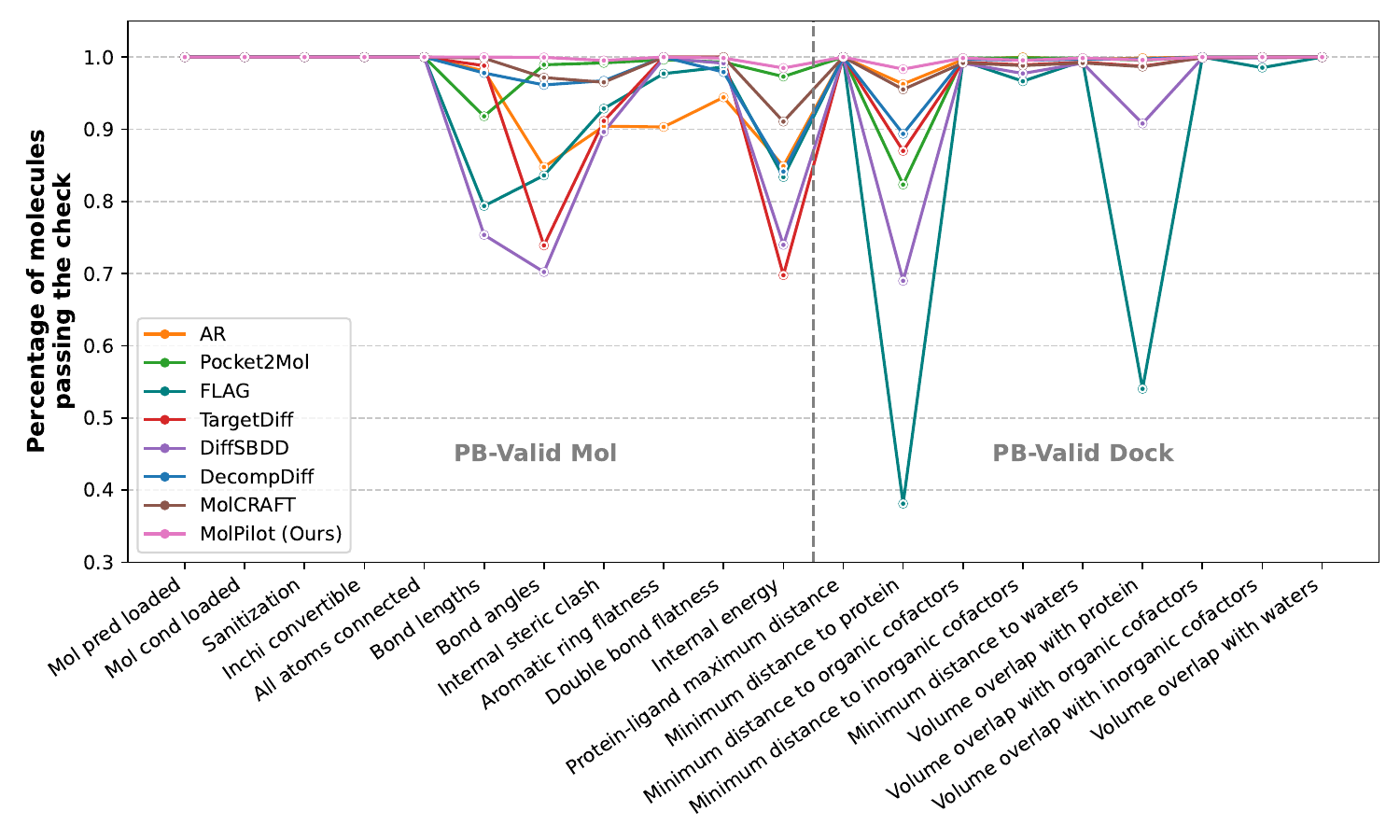}
    \caption{Percentage of generated molecules that have passed the PoseBusters validity checks on ID CrossDock test set. \emph{PB-Valid Mol:} intramolecular validity. \emph{PB-Valid Dock:} intermolecular validity.
    Reported \emph{PB-Valid}: {PB-Valid Mol} \& {PB-Valid Dock}.
    }
    \label{fig:pb_detail_id}
\end{figure*}

\begin{figure}[t]
    \centering
    \includegraphics[width=\linewidth]{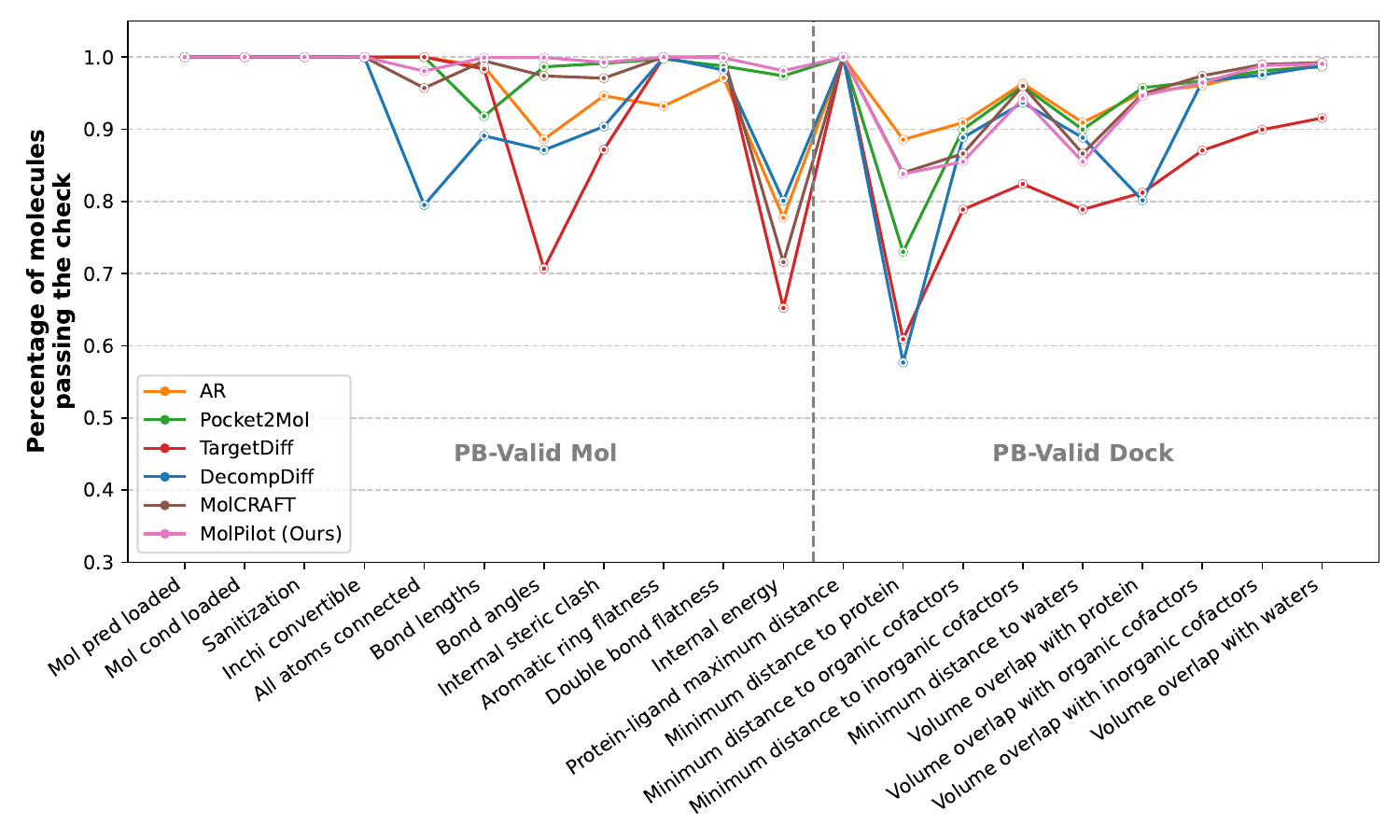}
    \caption{Percentage of generated molecules that have passed the PoseBusters validity checks on OOD PoseBusters test set. \emph{PB-Valid Mol:} intramolecular validity. \emph{PB-Valid Dock:} intermolecular validity.
        Reported \emph{PB-Valid}: PB-Valid Mol \& PB-Valid Dock.}
    \label{fig:pb_detail_ood}
\end{figure}

\subsection{Dynamics of Better-balanced Modalities}
\begin{figure}
    \centering
    \includegraphics[width=0.8\linewidth]{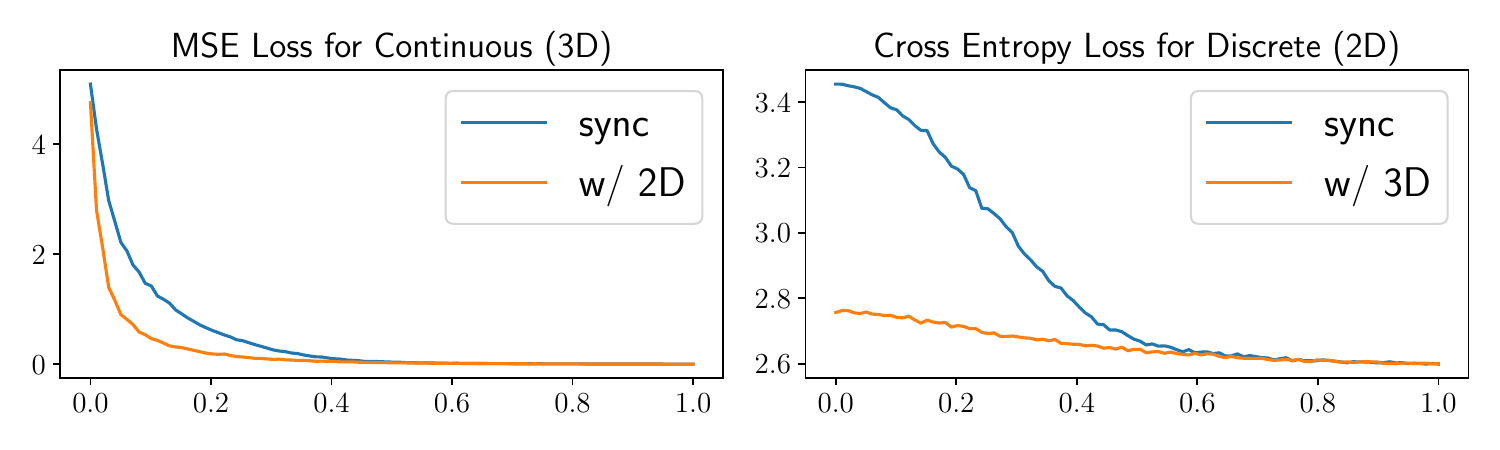}
    \caption{Validation loss curves for the model trained under generalized objective w.r.t. timestep in the generative process. \emph{Sync:} Modalities at the same timestep. \emph{w/ 2D:} Discrete modality always at $t=1$. \emph{w/ 3D:} Continuous modality always at $t=1$.}
    \label{fig:loss-decouple}
\end{figure}

We visualize the modality-specific validation loss curves for the model trained under generalized loss (Eq.~\ref{eq:generalized_loss}) in Fig.~\ref{fig:loss-decouple}, which shows that the denoising model effectively learns to mutually benefit from cleaner information either modality.

\begin{table}[tbp]
\caption{Results of Jenson-Shannon Divergence (JSD) of bond length, bond angle, torsion angle distributions between generated molecules and CrossDock reference molecules, averaged over all types with a frequency $>$ 100.}
\label{tab:jsd_results}
\begin{center}
\resizebox{0.5\linewidth}{!}{
\scriptsize
\begin{tabular}{l|c|c|c}
\toprule
\multirow{2}{*}{Methods} & Length ($\downarrow$) & Angle ($\downarrow$) & Torsion ($\downarrow$)\\
                         & Avg. JSD                    & Avg. JSD                   & Avg. JSD          \\ \midrule
AR         & 0.544          & 0.507          & 0.545          \\
Pocket2Mol & 0.472          & 0.482          & 0.467          \\
TargetDiff & 0.365          & 0.435          & 0.411          \\ 
DecompDiff & 0.332          & 0.410          & 0.338          \\
MolCRAFT   & 0.318          & 0.384          & 0.322          \\
Ours       & \textbf{0.252} & \textbf{0.351} & \textbf{0.287} \\
\bottomrule
\end{tabular}
}
\end{center}
\end{table}

\subsection{In-Distribution De novo Design}\label{subsec:app-exp-id}
We report the results on CrossDock \citep{francoeur2020three} in an in-distribution (ID) setting. We sample 100 molecules for each of the 100 test proteins, and the results in Table~\ref{tab:main} show that our model consistently performs exceptionally across most metrics—leading in PB-Valid, overall Vina affinities, and achieving comparable drug-like properties w.r.t. reference molecules.

For conformation quality, our model achieves the best performance with the highest PoseBusters passing rate, closely matching the reference value of 95.0\%. Details of each validity check are shown in Fig.~\ref{fig:pb_detail_id}, and it can be seen that the leading factors affecting the overall performance are bond lengths and angles (AR, TargetDiff), internal energy (MolCRAFT, DecompDiff) for intramolecular validity, and minimum distance to protein (Pocket2Mol) for intermolecular validity.

\begin{figure*}[h]
\centering
\subfigure[C-C Bond]{\includegraphics[width=\textwidth]{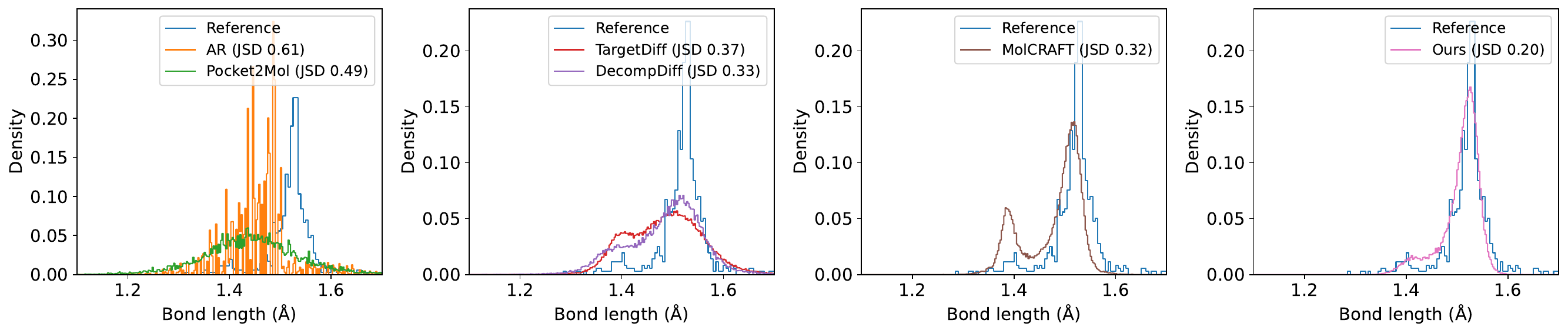}}
\subfigure[C:C Bond]{\includegraphics[width=\textwidth]{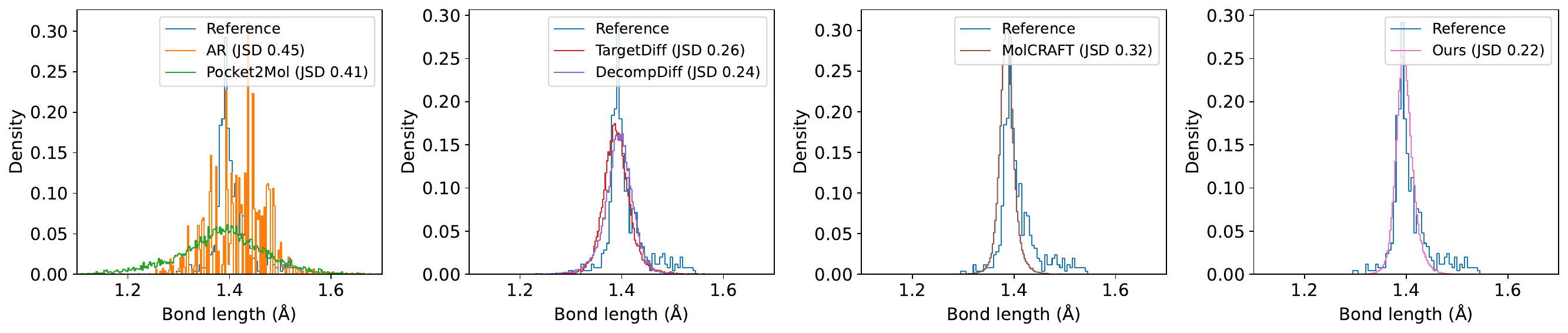}}
\subfigure[C-O Bond]{\includegraphics[width=\textwidth]{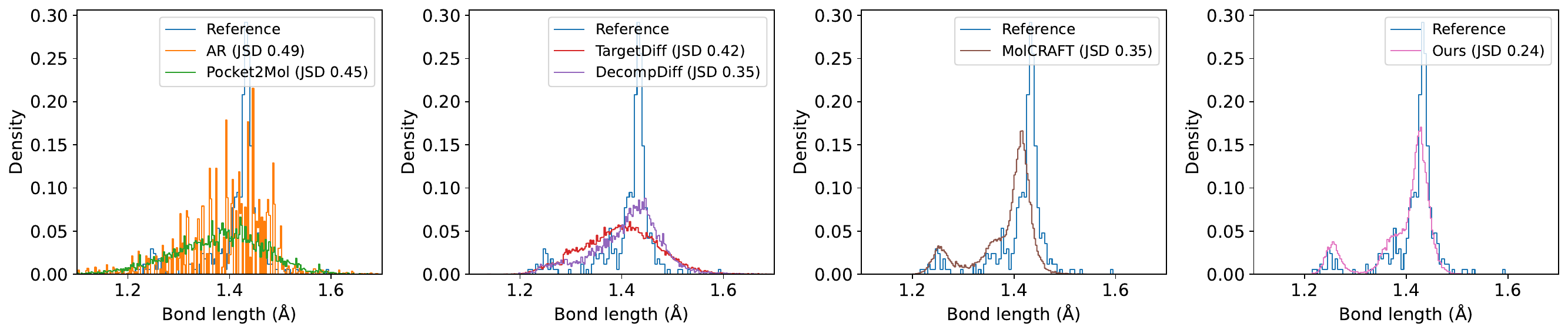}}
\subfigure[C-N Bond]{\includegraphics[width=\textwidth]{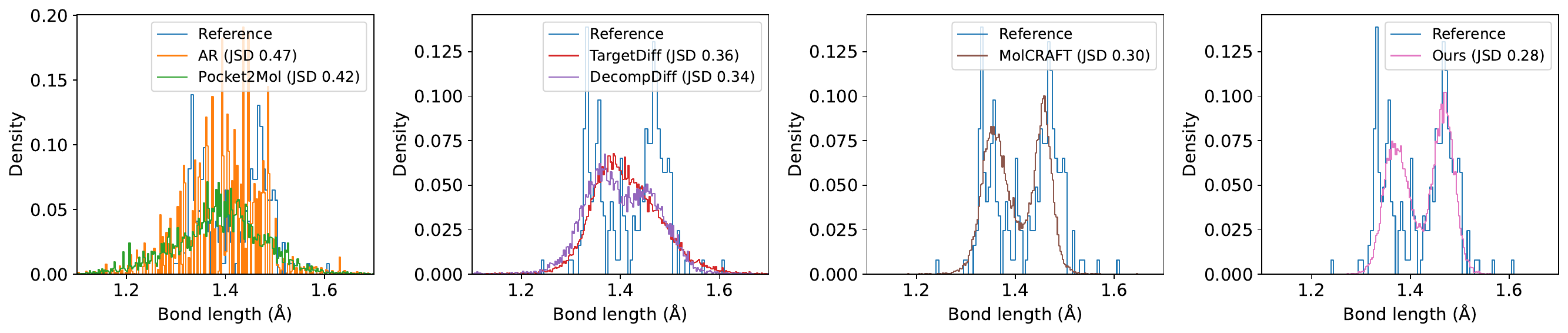}}
\subfigure[C:N Bond]{\includegraphics[width=\textwidth]{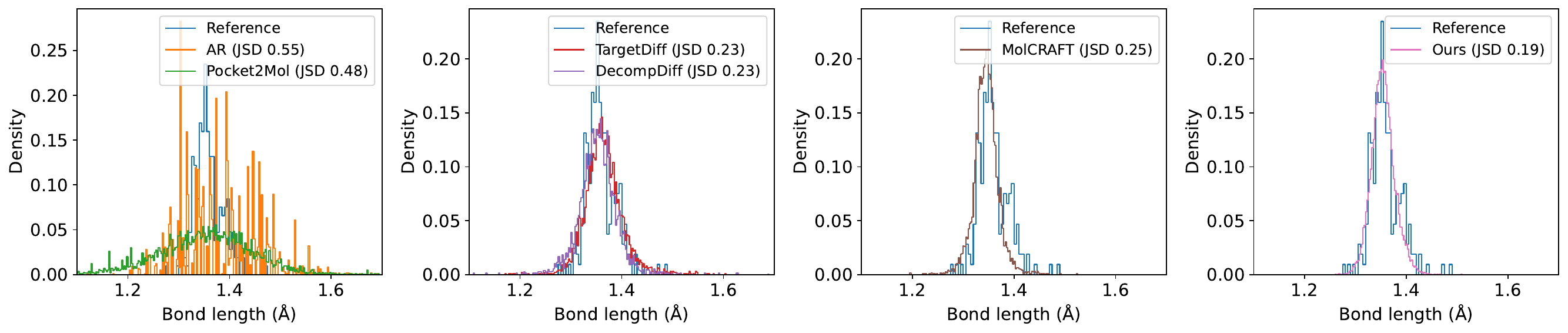}}
\caption{Top-5 frequent bond length distribution of generated molecules compared with CrossDock reference molecules.}
\label{fig:other_len}
\end{figure*}

\begin{figure*}[h]
\centering
\subfigure[C-C-C Bond Angle]{\includegraphics[width=\textwidth]{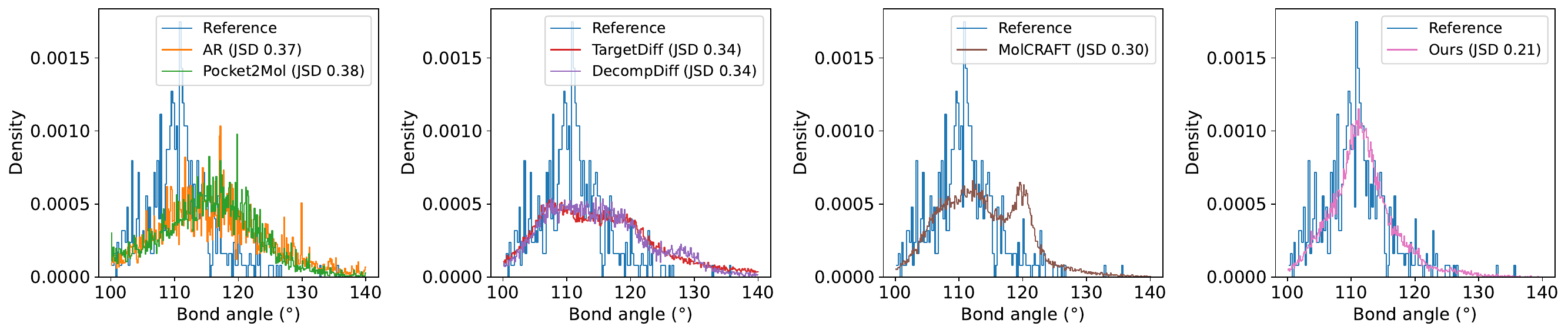}}
\subfigure[C:C:C Bond Angle]{\includegraphics[width=\textwidth]{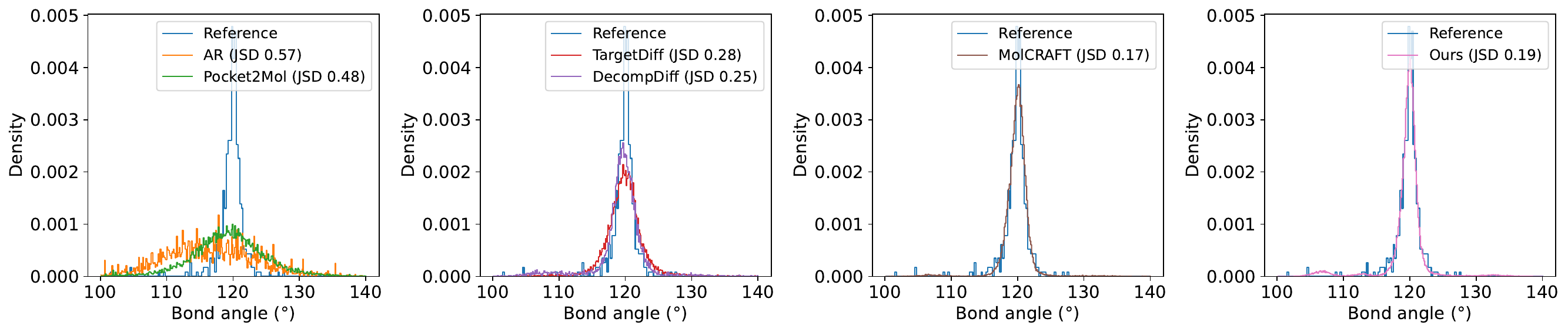}}
\subfigure[C-C-O Bond Angle]{\includegraphics[width=\textwidth]{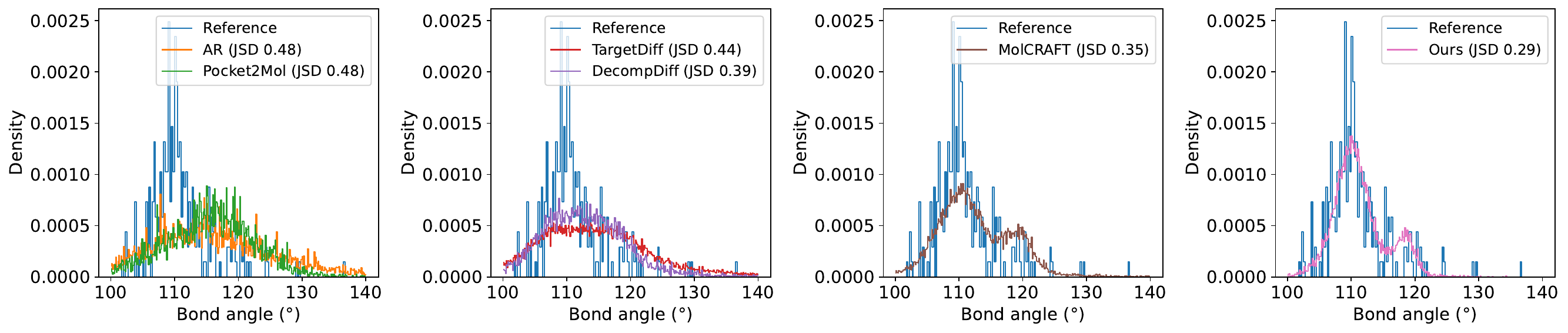}}
\subfigure[C:C:N Bond Angle]{\includegraphics[width=\textwidth]{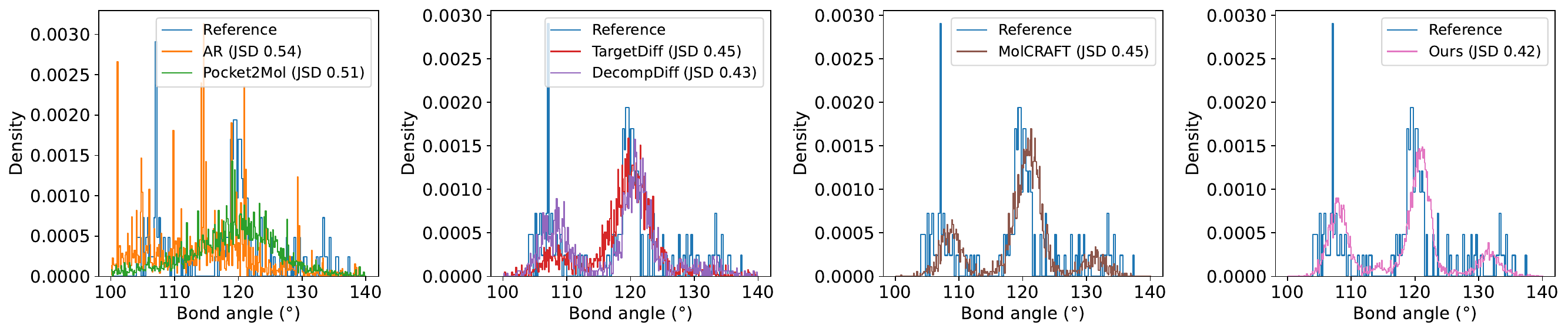}}
\subfigure[C-C-N Bond Angle]{\includegraphics[width=\textwidth]{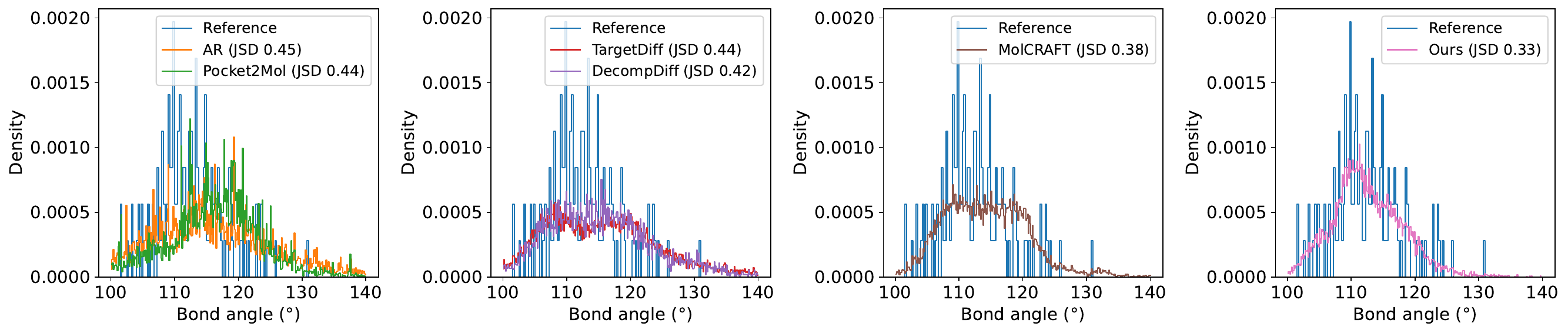}}
\caption{Top-5 frequent bond angle distribution of generated molecules compared with CrossDock reference molecules.}
\label{fig:other_angle}
\end{figure*}

\begin{figure*}[ht]
\centering
\subfigure[C-C-C-C Torsion Angle]{\includegraphics[width=\textwidth]{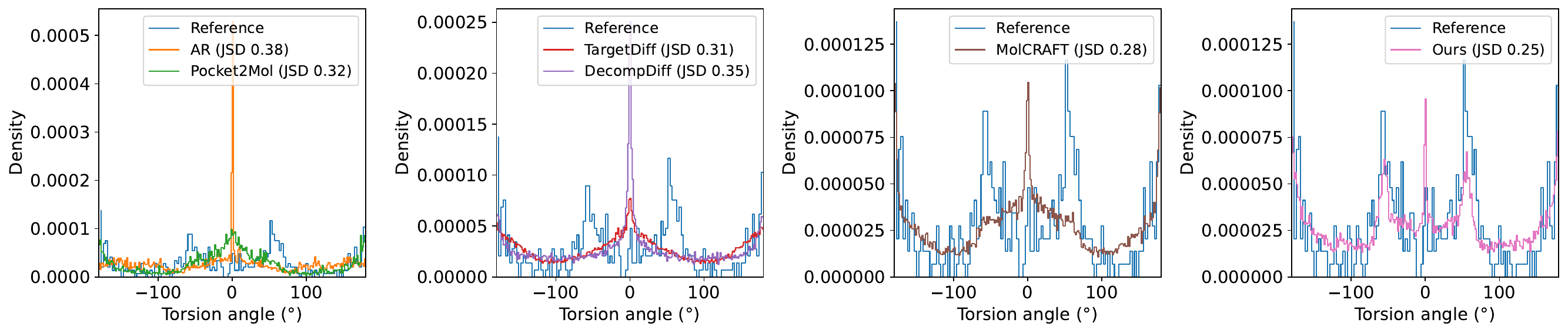}}
\subfigure[C:C:C:C Torsion Angle]{\includegraphics[width=\textwidth]{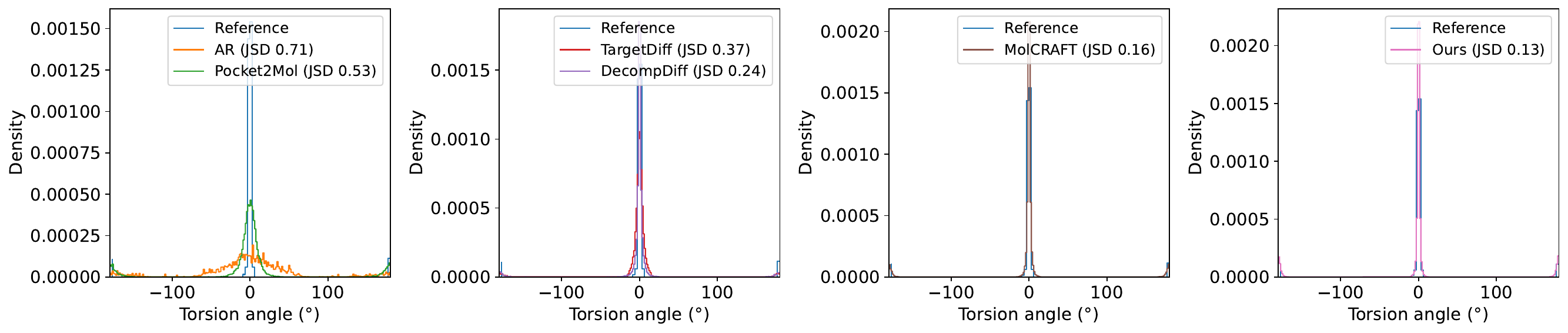}}
\subfigure[C-C-O-C Torsion Angle]{\includegraphics[width=\textwidth]{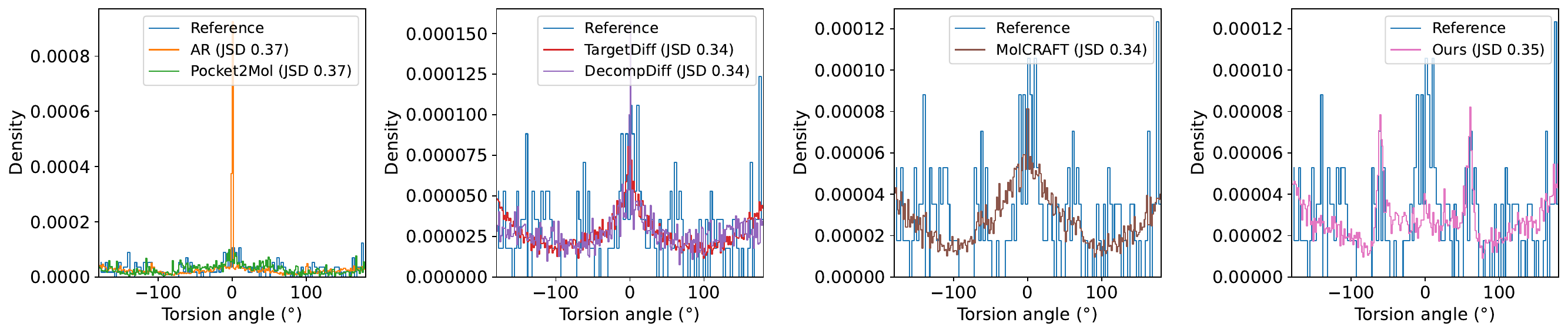}}
\subfigure[C-C-C-O Torsion Angle]{\includegraphics[width=\textwidth]{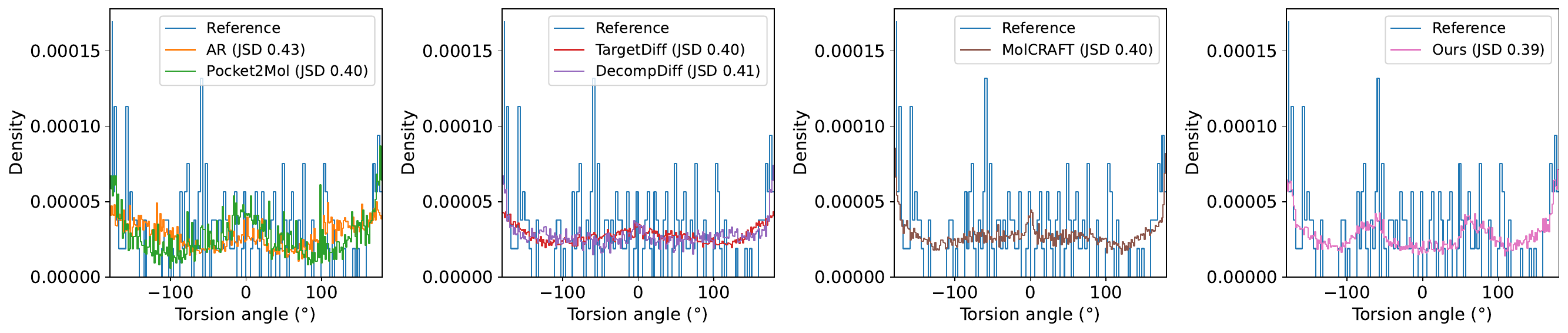}}
\subfigure[C-C-N-C Torsion Angle]{\includegraphics[width=\textwidth]{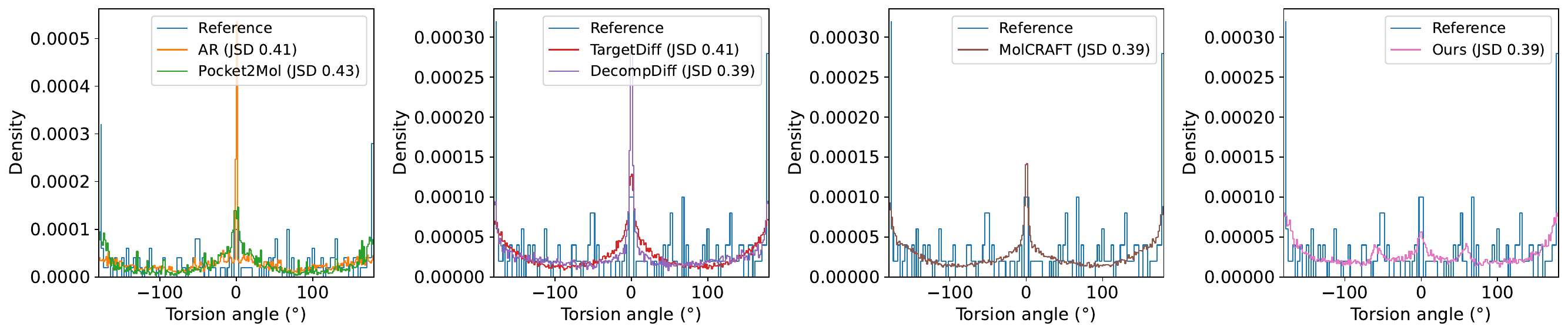}}
\caption{Top-5 frequent torsion angle distribution of generated molecules compared with CrossDock reference molecules.}
\label{fig:other_torsion}
\end{figure*}

For molecular geometries, we visualize the distributions of bond length, bond angle, and torsion angle, between generated molecules and reference molecules in the test set in Fig.~\ref{fig:other_len}, \ref{fig:other_angle}, \ref{fig:other_torsion} for the top-5 frequent types, and summarize the overall Jensen-Shannon Divergence (JSD) in Table~\ref{tab:jsd_results}, averaged over all types with a frequency $> 100$. While the previous strong-performing method MolCRAFT captures the bond length distributions relatively well, it cannot fit all the bond angle or torsion angle distributions, displaying for example non-standard C-C-C bond angle, smoothed C-C-N bond angle, and diverges more severely in C-C-C-C and C-C-N-C torsion angles compared with reference. Our method excels at modeling all these distributions, underscoring its ability to capture the molecular geometries.

For binding affinities, our model demonstrates the leading performance with an average Vina score of -6.88 kcal/mol, maintaining the closest gap between Vina Score, Vina Min, and Vina Dock. This shows our model's superiority in capturing spatial interactions in the generated pose.

For molecular properties like QED and SA, our model displays competitive performance, and ranks at the top among all non-autoregressive models.

\subsection{Out-of-Distribution De novo Design}
\begin{table}[htbp]
\centering
\caption{Strain Energy results calculated by PoseCheck v1.1, and PoseBusters internal energy passing rate for SBDD baselines.}
\label{tab:strain_results}
\vskip 0.1in
\resizebox{0.35\columnwidth}{!}{%
\begin{tabular}{@{}l|cccc@{}}
\toprule
\multirow{2}{*}{Methods} & \multicolumn{4}{c}{Strain Energy}                                                        \\
                         & 25\%               & 50\%         & {75\%}          & Passed          \\ \midrule
Reference                & 2.6                & 7.1          & {18.9}          & 100\%           \\ \midrule
AR                       & 5.6                & 40.1         & {238.8}         & 77.7\%          \\
Pocket2Mol               & \textbf{1.0}       & \textbf{4.4} & {\textbf{28.5}} & \ul{97.4\%}    \\ \midrule
TargetDiff               & 41.4               & 158.2        & {625.1}         & 65.2\%          \\
DecompDiff               & 6.8                & 41.5         & {249.2}         & 80.1\%          \\
MolCRAFT                 & 5.3                & 46.5         & {27397.8}       & 71.6\%          \\ \midrule
Ours                     & \ul{{2.7}} & \ul{10.6}   & {\ul{38.1}}    & \textbf{98.1\%} \\ \bottomrule
\end{tabular}%
}
\end{table}

\begin{figure}
    \centering
    \includegraphics[width=\linewidth]{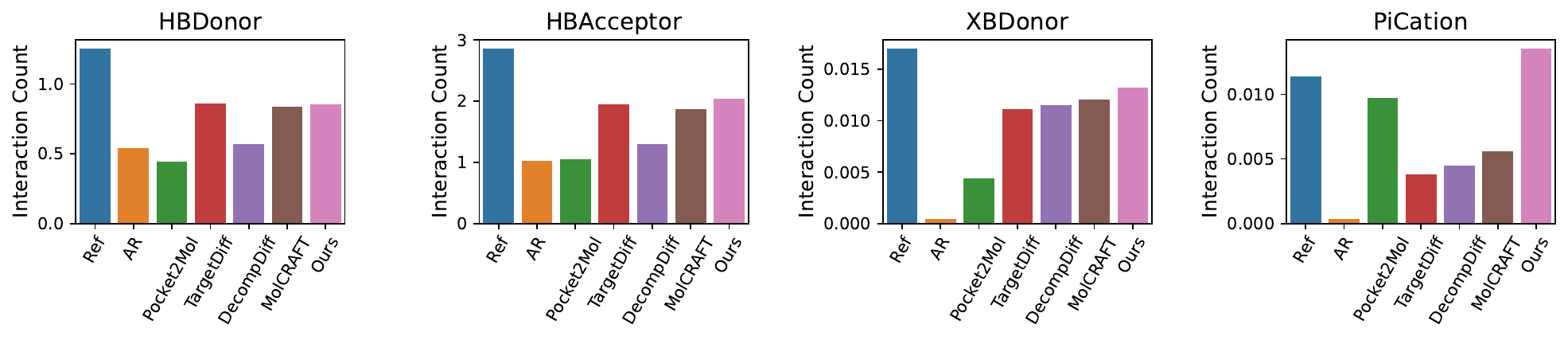}
    \caption{Detailed interaction counts for generated molecules on PoseBusters test. Only PB-Valid molecules are considered.}
    \label{fig:interaction_normalized}
\end{figure}

We select competitive baselines in the ID settings, and evaluate them on the OOD PoseBusters test set. We sample 100 molecules for each of the 180 test proteins, and report the overall results in Table~\ref{tab:main}.

We provide more detailed statistics of PoseBusters passing rate in Fig.~\ref{fig:pb_detail_ood}. It can be seen that the performance drop in PB-Valid is mainly attributed to two major factors: (1) \emph{Intramolecular validity} drops due to distorted molecular geometry such as non-standard bond lengths, bond angles, ring flatness and the most prominently, internal energy as can be viewed as the overall indicator of molecular geometries, where all the non-autoregressive baselines and AR are around or below 80\%. (2) \emph{Intermolecular validity} drops due to clahses w.r.t. protein surface. TargetDiff appears to be mostly affected, and nearly all models cannot effectively avoid clashing by maintaining a feasible distance to protein atoms, waters or other cofactors. 
For protein atom clashing, it is worth pointing out that the high performance over CrossDock might come from the potential information leakage caused by data splits (Appendix~\ref{subsec:app-problem-cd}). Thus the passing rates on held-out PoseBusters test set can serve as a more informative indicator of the performances between different methods, showing that voxel-based AR and BFN-based models are performing better than diffusion-based models. However, it should be noted that though with better clashing performance, AR generates significantly fewer molecules that are also with smaller sizes, undermining its practicality. 
Besides, current SBDD models have not yet taken waters and cofactors into consideration, suggesting that there might be a calling for an advanced and comprehensive formulation of SBDD.

For a finer-grained quantification of the internal energy, we employ PoseCheck test suite \citep{harris2023benchmarking} to measure the Strain Energy of generated molecules, and report the percentiles of distributions in Table~\ref{tab:strain_results}, associated with the internal energy passing rate in PoseBusters checks. It shows that MolCRAFT generates considerably more strained structures as suggested by the tail distribution, and our \ours~and Pocket2Mol maintains the robust performance.

We provide the detailed interaction counts compared with ground-truth complex structures in PoseBusters. Fig.~\ref{fig:interaction_normalized} shows that \ours~matches the genuine interaction profiles relatively well, especially in Cation-$\pi$ interactions. We additionally report in Table~\ref{tab:interaction_sim_full} the full statistics of interaction profile similarity calculated from different poses.
It can be seen that our \ours~and MolCRAFT closely match the interaction fingerprint, but MolCRAFT suffers from strained poses, resulting in unrealistic interactions for generated poses with high internal energy. In contrast, our model is able to capture the true interactions reliably.

\begin{table}[htbp]
\caption{Tanimoto similarity of the interaction profiles between ground-truth structures in PoseBusters and molecules generated by SBDD models. \emph{Gen:} interactions directly calculated from generated poses. \emph{Redock:} interactions calculated from Vina redocked poses. \emph{Gen \& PB-Valid:} interactions directly calculated from generated poses that have passed the PoseBusters validity checks.}
\label{tab:interaction_sim_full}
\centering
\vskip 0.1in
\resizebox{0.55\linewidth}{!}{%
\begin{tabular}{@{}lccc@{}}
\toprule
Methods    & Sim. (Gen)     & Sim. (Redock)  & Sim. (Gen \& PB-Valid) \\ \midrule
AR         & 0.394          & 0.344          & 0.221                  \\
Pocket2Mol & 0.447          & 0.405          & 0.436                  \\
TargetDiff & 0.477          & 0.415          & 0.458                  \\
DecompDiff & 0.438          & 0.332          & 0.374                  \\
MolCRAFT   & \textbf{0.560} & \ul{0.444}    & \ul{0.498}            \\
Ours       & \ul{0.552}    & \textbf{0.477} & \textbf{0.551}         \\ \bottomrule
\end{tabular}%
}
\end{table}

\begin{figure*}[htbp]
    \centering
    \includegraphics[width=\linewidth]{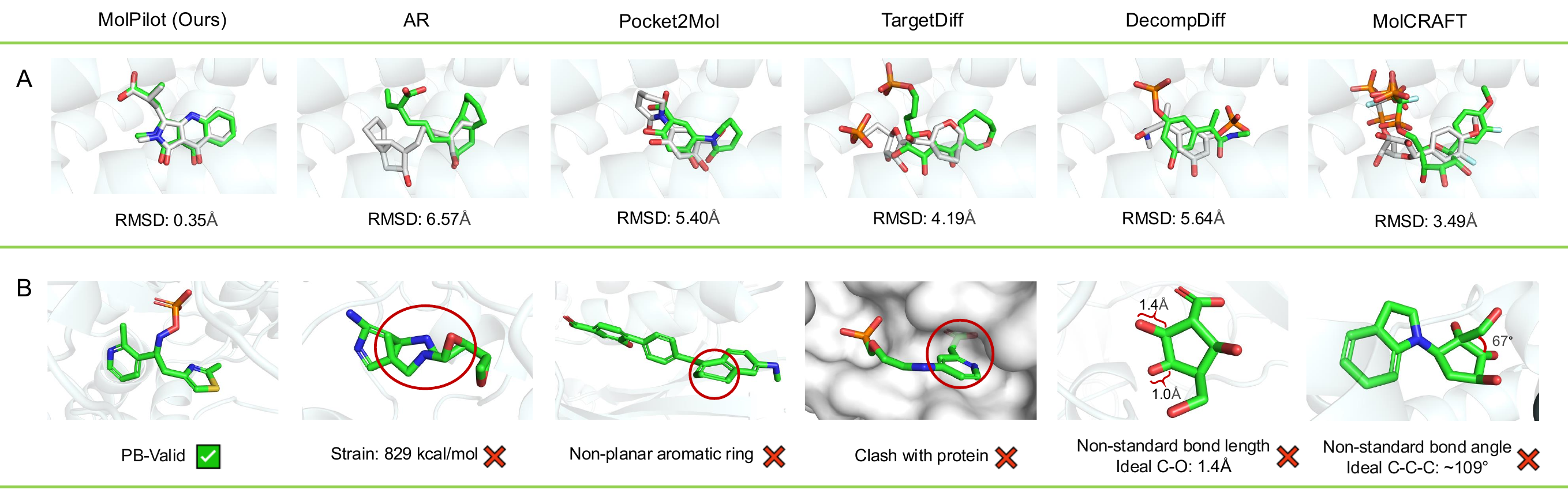}
    \caption{Illustration of molecular geometries for SBDD models. \textbf{A.} Binding mode consistency upon redocking, measured by scRMSD between generated poses (\textcolor{mygreen}{Green}) and redocked poses ({\color{gray}Silver}). \textbf{B.} Generated molecules with unrealistic geometries, thus failing to pass PoseBusters validity checks.}
    \label{fig:baseline_fail}
\end{figure*}

\subsection{Molecular Docking}
We repurpose the SBDD models from the joint codesign model $P(\vx, \vh, \mA \mid \rvx_P)$ to the conditional marginal $P(\vx \mid \vh, \mA, \rvx_P)$. This actually corresponds to another specific line on the loss surface described by Eq.~\ref{eq:VLB}, with $\beta_d(t) \equiv \beta_d(1)$. We employ the joint schedule $\beta(t)=(\tilde{\beta}_c(t), \tilde{\beta}_d(1))$ with $\tilde{\beta}$ being the default noise schedule for continuous modality in Eq.~\ref{eq:beta}.

\end{document}